\input harvmac
\input epsf

%
\let\includefigures=\iftrue
%
%
%
\newfam\black
\input rotate
\input epsf
\noblackbox
%
%
\includefigures
\message{If you do not have epsf.tex (to include figures),}
\message{change the option at the top of the tex file.}
\def\figin{\epsfcheck\figin}\def\figins{\epsfcheck\figins}
\def\epsfcheck{\ifx\epsfbox\UnDeFiNeD
\message{(NO epsf.tex, FIGURES WILL BE IGNORED)}
\gdef\figin##1{\vskip2in}\gdef\figins##1{\hskip.5in}
\else\message{(FIGURES WILL BE INCLUDED)}%
\gdef\figin##1{##1}\gdef\figins##1{##1}\fi}
\def\DefWarn#1{}
\def\N{{\cal N}}
\def\figinsert{\goodbreak\midinsert}
\def\ifig#1#2#3{\DefWarn#1\xdef#1{fig.~\the\figno}
\writedef{#1\leftbracket fig.\noexpand~\the\figno}%
\figinsert\figin{\centerline{#3}}\medskip\centerline{\vbox{\baselineskip12pt
\advance\hsize by -1truein\noindent\footnotefont{\bf
Fig.~\the\figno:} #2}}
\bigskip\endinsert\global\advance\figno by1}
\else
\def\ifig#1#2#3{\xdef#1{fig.~\the\figno}
\writedef{#1\leftbracket fig.\noexpand~\the\figno}%
\global\advance\figno by1} \fi

\def\tilde{\widetilde}

\def\yboxit#1#2{\vbox{\hrule height #1 \hbox{\vrule width #1
\vbox{#2}\vrule width #1 }\hrule height #1 }}
\def\fillbox#1{\hbox to #1{\vbox to #1{\vfil}\hfil}}
\def\ybox{{\lower 1.3pt \yboxit{0.4pt}{\fillbox{8pt}}\hskip-0.2pt}}

\def\l{\left}

\def\rightarrowbox#1#2{
  \setbox1=\hbox{\kern#1{${ #2}$}\kern#1}
  \,\vbox{\offinterlineskip\hbox to\wd1{\hfil\copy1\hfil}
    \kern 3pt\hbox to\wd1{\rightarrowfill}}}

\def\Tr{{{\rm Tr~ }}}

\def\vev#1{\langle{#1}\rangle}

\def\tilde{\widetilde}

\def\II{\relax{I\kern-.10em I}}

\def\bar{\overline}

\def\IZ{\relax\ifmmode\mathchoice
{\hbox{\cmss Z\kern-.4em Z}}{\hbox{\cmss Z\kern-.4em Z}}
{\lower.9pt\hbox{\cmsss Z\kern-.4em Z}} {\lower1.2pt\hbox{\cmsss
Z\kern-.4em Z}}\else{\cmss Z\kern-.4em Z}\fi}
\def\IB{\relax{\rm I\kern-.18em B}}
\def\IC{{\relax\hbox{$\inbar\kern-.3em{\rm C}$}}}
\def\ID{\relax{\rm I\kern-.18em D}}
\def\IE{\relax{\rm I\kern-.18em E}}
\def\IF{\relax{\rm I\kern-.18em F}}
\def\IG{\relax\hbox{$\inbar\kern-.3em{\rm G}$}}
\def\IGa{\relax\hbox{${\rm I}\kern-.18em\Gamma$}}
\def\IH{\relax{\rm I\kern-.18em H}}
\def\II{\relax{\rm I\kern-.18em I}}
\def\IK{\relax{\rm I\kern-.18em K}}
\def\IN{\relax{\rm I\kern-.18em N}}
\def\IP{\relax{\rm I\kern-.18em P}}

%
\def\inbar{\,\vrule height1.5ex width.4pt depth0pt}

\font\cmss=cmss10 \font\cmsss=cmss10 at 7pt
\def\IR{\relax{\rm I\kern-.18em R}}

\def\lp10{l_P^{10}}
\def\lp11{l_P^{11}}
\def\R11{R_{11}}

\def\l{\lambda}
\def\lt{\tilde\lambda}

\def\gb#1{ {\langle #1 ] } }

\newbox\tmpbox\setbox\tmpbox\hbox{\abstractfont}
\Title{\vbox{\baselineskip12pt\hbox to\wd\tmpbox{\hss
 hep-th/0506126} }}
{\vbox{\centerline{Two-Loop Amplitudes of Gluons and Octa-Cuts  }
\bigskip
\centerline{in $\N=4$ Super Yang-Mills}
 }}
\smallskip
\centerline{Evgeny I. Buchbinder$^a$ and Freddy Cachazo$^b$}
\smallskip
\bigskip
\centerline{\it $^a$ School of Natural Sciences, Institute for
Advanced Study, Princeton NJ 08540 USA}
\bigskip
\centerline{\it $^b$ Perimeter Institute for Theoretical Physics,
Waterloo, Ontario N2J 2W9, Canada}

\bigskip
\vskip 1cm \noindent

\input amssym.tex

After reduction techniques, two-loop amplitudes in ${\cal N}=4$
super Yang-Mills theory can be written in a basis of integrals
containing scalar double-box integrals with rational coefficients,
though the complete basis is unknown. Generically, at two loops,
the leading singular behavior of a scalar double box integral with
seven propagators is captured by a hepta-cut. However, it turns
out that a certain class of such integrals has an additional
propagator-like singularity. One can then formally cut the new
propagator to obtain an octa-cut which localizes the cut integral
just as a quadruple cut does at one-loop. This immediately gives
the coefficient of the scalar double box integral as a product of
six tree-level amplitudes. We compute, as examples, several
coefficients of the five- and six-gluon non-MHV two-loop
amplitudes. We also discuss possible generalizations to higher
loops.

\Date{June 2005}

\lref\WittenNN{
E.~Witten,
``Perturbative gauge theory as a string theory in twistor space'',
Commun. Math. Phys. 252, 189 (2004)
[hep-th/0312171].}

\lref\MHVdec{
F. Cachazo, P. Svr\v{c}ek and E. Witten,
``MHV vertices and tree amplitudes in gauge theory'',
JHEP 0409:006 (2004)
[hep-th/0403047].}

\lref\treeone{
C. J. Zhu,
``The googly amplitudes in gauge theory'',
JHEP {0404}:032 (2004)
[hep-th/0403115].}

\lref\treetwo{
G.\ Georgiou and V. V. Khoze,
``Tree amplitudes in gauge theory as scalar MHV diagrams'',
JHEP 0405:070 (2004)
[hep-th/0404072].}

\lref\treethree{
J. B. Wu and C. J. Zhu,
``MHV vertices and scattering amplitudes in gauge theory'',
JHEP 0407:032 (2004)
[hep-th/0406085].}

\lref\treefour{
J. B. Wu and C. J. Zhu,
``MHV vertices and fermionic scattering amplitudes in gauge theory with quarks and gluinos'',
JHEP 0409:063 (2004)
[hep-th/0406146].}

\lref\treefive{
D. A. Kosower,
``Next-to-maximal helicity violating amplitudes in gauge theory'',
Phys. Rev. D71:045007 (2005)
[hep-th/0406175].}

\lref\treesix{
G. Georgiou, E. W. N. Glover and V. V. Khoze,
``Non-MHV tree amplitudes in gauge theory'',
JHEP 0407:048 (2004)
[hep-th/0407027].}

\lref\treeseven{
Y. Abe, V. P. Nair and M. I. Park,
``Multigluon amplitudes, N = 4 constraints and the WZW model'',
Phys. Rev. D71:025002 (2005)
[hep-th/0408191].}

\lref\Higgs{
L.~J.~Dixon, E.~W.~N.~Glover and V.~V.~Khoze,
``MHV rules for Higgs plus multi-gluon amplitudes'',
JHEP 0412:015 (2004)
[hep-th/0411092].}

\lref\Higgsone{
S. D. Badger, E. W. N. Glover and V. V. Khoze,
``MHV rules for Higgs plus multi-parton amplitudes'',
JHEP 0503:023 (2005)
[hep-th/0412275].}

\lref\Currents{
Z. Bern, D. Forde, D. A. Kosower and P. Mastrolia,
``Twistor-inspired construction of electroweak vector boson currents'',
hep-ph/0412167.}

\lref\RSVNewTree{
R. Roiban, M. Spradlin and A. Volovich,
``Dissolving N = 4 loop amplitudes into QCD tree amplitudes'',
Phys. Rev. Lett. 94:102002 (2005)
[hep-th/0412265].}

\lref\recurone{
R. Britto, F. Cachazo and B. Feng,
``New recursion relations for tree amplitudes of gluons'',
Nucl. Phys. B715 (2005) 499-522
[hep-th/0412308].}

\lref\recurtwo{
R. Britto, F. Cachazo, B. Feng and E. Witten,
``Direct proof of tree-level recursion relation in Yang-Mills theory'',
hep-th/0501052.}

\lref\LuoWen{
M. Luo and C. Wen,
``Recursion relations for tree amplitudes in super gauge theories'',
JHEP 0503:004 (2005)
[hep-th/0501121].}

\lref\LuoWenone{
M. Luo and C. Wen,
``Compact formulas for all tree amplitudes of six partons'',
hep-th/0502009.}

\lref\BFRSV{
R. Britto, B. Feng, R. Roiban, M. Spradlin and A. Volovich,
``All split helicity tree-level gluon amplitudes'',
hep-th/0503198.}

\lref\BadgerMassive{
S. D. Badger, E. W. N. Glover, V. V. Khoze and P. Svr\v{c}ek,
``Recursion Relations for Gauge Theory Amplitudes with Massive Particles'',
hep-th/0504159.}

\lref\QCDone{
Z. Bern, L. J. Dixon and D. A. Kosower,
``On-Shell Recurrence Relations for One-Loop QCD Amplitudes'',
hep-th/0501240.}

\lref\QCDtwo{Z. Bern, L. J. Dixon and D. A. Kosower,
``The Last of the Finite Loop Amplitudes in QCD'',
hep-ph/0505055.}

\lref\BrandhuberJW{
A.~Brandhuber, S.~McNamara, B.~Spence and G.~Travaglini,
``Loop Amplitudes in Pure Yang-Mills from Generalised Unitarity,''
hep-th/0506068.}

\lref\OtherGaugeCalcs{
C. Quigley and M. Rozali,
``One-loop MHV amplitudes in supersymmetric gauge theories'',
JHEP 0501:053 (2005)
[hep-th/0410278].}

\lref\oneloopone{
J. Bedford, A. Brandhuber, B. Spence and G. Travaglini,
``A twistor approach to one-loop amplitudes in N = 1 supersymmetric Yang-Mills
theory'',
Nucl. Phys. B706:100 (2005)
[hep-th/0410280].}

\lref\onelooptwo{
J. Bedford, A.~Brandhuber, B.~Spence and G.~Travaglini,
``Non-Supersymmetric Loop Amplitudes and MHV Vertices'',
Nucl. Phys. B712:59 (2005)
[hep-th/0412108].}

\lref\oneloopthree{
S. J. Bidder, N. E. J. Bjerrum-Bohr, L. J. Dixon and D. C. Dunbar,
``N = 1 supersymmetric one-loop amplitudes and the holomorphic anomaly of
unitarity cuts'',
Phys. Lett. B606:189 (2005)
[hep-th/0410296].}

\lref\oneloopfour{
S. J. Bidder, N. E. J. Bjerrum-Bohr, D. C. Dunbar and W. B. Perkins,
``Twistor space structure of the box coefficients of N = 1 one-loop
amplitudes'',
Phys. Lett. B608:151 (2005)
[hep-th/0412023].}

\lref\oneloopfive{
S. J. Bidder, N. E. J. Bjerrum-Bohr, D. C. Dunbar and W. B. Perkins,
``One-loop gluon scattering amplitudes in theories with $N < 4$
supersymmetries'',
Phys. Lett. B612:75 (2005)
[hep-th/0502028].}

\lref\oneloopus{
R. Britto, E. I. Buchbinder, F. Cachazo and B. Feng,
``One-loop amplitudes of gluons in SQCD'',
hep-ph/0503132.}

\lref\BST{
A. Brandhuber, B. Spence and G. Travaglini,
``One-loop gauge theory amplitudes in N = 4 super Yang-Mills from MHV
vertices'',
Nucl. Phys. B706:150 (2005)
[hep-th/0407214].}

\lref\CachazoAnomaly{
F. Cachazo,
``Holomorphic Anomaly Of Unitarity Cuts And One-Loop Gauge Theory
Amplitudes'',
hep-th/0410077.}

\lref\BCFseven{
R. Britto, F. Cachazo and B. Feng,
``Computing one-loop amplitudes from the holomorphic anomaly of unitarity
cuts'',
Phys. Rev. D71:025012 (2005)
[hep-th/0410179].}

\lref\NeqFourSevenPoint{
Z. Bern, V. Del Duca, L. J. Dixon and D. A. Kosower,
``All non-maximally-helicity-violating one-loop seven-gluon amplitudes
in N = 4 super-Yang-Mills theory'',
Phys. Rev. D71:045006 (2005)
[hep-th/0410224].}

\lref\General{
R. Britto, F. Cachazo and B. Feng,
``Generalized Unitarity and One-Loop Amplitudes in N=4 Super-Yang-Mills'',
hep-th/0412103.}

\lref\NeqFourNMHV{
Z. Bern, L. J. Dixon and D. A. Kosower,
``All next-to-maximally helicity-violating one-loop gluon amplitudes in N = 4
super-Yang-Mills theory'',
hep-th/0412210.}

\lref\DixonWI{ L.~J.~Dixon, ``Calculating Scattering Amplitudes
Efficiently,'' hep-ph/9601359.}

\lref\conjecture{C. Anastasiou, Z. Bern, L. Dixon and D. A. Kosower,
``Planar Amplitudes in Maximally Supersymmetric Yang-Mills Theory'',
Phys.Rev.Lett. 91 (2003) 251602 [hep-th/0309040].}

\lref\hundred{Z. Bern, L. J. Dixon and D. A. Kosower,
``Two-Loop $g \to gg$ Splitting Amplitudes in QCD'',
JHEP 0408 (2004) 012 [hep-ph/0404293].}

\lref\Kosower{D. A. Kosower,
``All-Order Collinear Behavior in Gauge Theories'',
Nucl.Phys. B552 (1999) 319-336 [hep-ph/9901201].}

\lref\high{C. Anastasiou, L.~J.~Dixon, Z. Bern and D.~A.~Kosower,
``Cross-Order Relations in $N=4$ Supersymmetric Gauge Theory,"
hep-th/0402053}

\lref\New{Z. Bern, L. J. Dixon and V. A. Smirnov,
``Iteration of Planar Amplitudes in Maximally Supersymmetric Yang-Mills Theory at Three Loops and Beyond,''
hep-th/0505205}

\lref\passarino{ L.~M. Brown and R.~P. Feynman, ``Radiative
Corrections To Compton Scattering,'' Phys. Rev. 85:231 (1952);
G.~Passarino and M.~Veltman, ``One Loop Corrections For E+ E-
Annihilation Into Mu+ Mu- In The Weinberg Model,'' Nucl. Phys.
B160:151 (1979); G.~'t Hooft and M.~Veltman, ``Scalar One Loop
Integrals,'' Nucl. Phys. B153:365 (1979); R.~G.~ Stuart,
``Algebraic Reduction Of One Loop Feynman Diagrams To Scalar
Integrals,'' Comp. Phys. Comm. 48:367 (1988); R.~G.~Stuart and
A.~Gongora, ``Algebraic Reduction Of One Loop Feynman Diagrams To
Scalar Integrals. 2,'' Comp. Phys. Comm. 56:337 (1990).}

\lref\thebook{R. J. Eden, P. V. Landshoff, D. I. Olive and J. C. Polkinghorne,
{\it The Analytic S-Matrix}, Cambridge University Press, 1966.}

\lref\Bernry{Z. Bern, J.S. Rozowsky and B. Yan,
``Two-Loop Four-Gluon Amplitudes in N=4 Super-Yang-Mills'',
Phys.Lett. B401 (1997) 273-282 [hep-ph/9702424].}

\lref\smirnov{V. A. Smirnov, Phys. Lett. B {\bf 460}, 397 (1999)
[hep-ph/9905323]; V.~A.~Smirnov,
  ``Analytical Result for Dimensionally Regularized Massless Master Double Box
  With One Leg Off Shell,''
  Phys.\ Lett.\ B {\bf 491}, 130 (2000)
  [arXiv:hep-ph/0007032];V.~A.~Smirnov,
  ``Analytical Result for Dimensionally Regularized Massless Master Non-planar
  Double Box With One Leg Off Shell,''
  Phys.\ Lett.\ B {\bf 500}, 330 (2001)
  [arXiv:hep-ph/0011056].
}

\lref\berends{F. A. Berends, W. T. Giele and H. Kuijf, ``On
Relations Between Multi-Gluon And Multi-Graviton Scattering,"
Phys. Lett B211 (1988) 91.}

\lref\xu{Z. Xu, D.-H. Zhang and L. Chang, ``Helicity Amplitudes For Multiple
Bremsstrahlung In Massless Nonabelian Theories,''
Nucl. Phys. {\bf B291}
(1987) 392.}

\lref\gunion{J.~F. Gunion and Z. Kunszt,
``Improved Analytic Techniques For Tree Graph Calculations
And The G G Q
Anti-Q Lepton Anti-Lepton Subprocess'',
Phys. Lett. 161B
(1985) 333.}

\lref\colorordone{Z. Bern and D. A. Kosower,
``Color Decomposition Of One Loop Amplitudes In Gauge Theories,''
Nucl.\ Phys. B362, 389 (1991)}

\lref\colorordtwo{F. A. Berends and W. Giele,
``The Six Gluon Process As An Example Of Weyl-Van Der Waerden Spinor Calculus,''
Nucl. Phys. B294, 700 (1987)}

\lref\colorordthree{M. Mangano, S. J. Parke and Z. Xu,
``Duality And Multi - Gluon Scattering,''
Nucl. Phys. B298 (1988) 653}

\lref\colorordfour{M.~L.~Mangano,
``The Color Structure Of Gluon Emission,''
Nucl. Phys. B309, 461 (1988)}

\lref\cape{F. Cachazo and P. Svrcek, ``Lectures on Twistor Strings
and Perturbative Yang-Mills Theory,'' hep-th/0504194}

\lref\mangparke{M. Mangano and S.~J. Parke, ``Multiparton Amplitudes In Gauge Theories,''
Phys. Rep. 200
(1991) 301.}

\lref\BernZX{ Z. Bern, L. J. Dixon, D. C. Dunbar and
D. A. Kosower, ``One Loop N Point Gauge Theory Amplitudes,
Unitarity And Collinear Limits,'' Nucl. Phys. B425, 217
(1994) [hep-ph/9403226].}

\lref\Fus{Z. Bern, L. Dixon, D.C. Dunbar and D.A. Kosower,
``Fusing Gauge Theory Tree Amplitudes Into Loop Amplitudes,''
Nucl.Phys. B435 (1995) 59-101 [hep-ph/9409265].}

\lref\BernDB{
Z.~Bern and A.~G.~Morgan,
``Massive Loop Amplitudes from Unitarity,''
Nucl.\ Phys.\ B 467, 479 (1996)
[hep-ph/9511336].}

\lref\BernJE{
Z.~Bern, L.~J.~Dixon and D.~A.~Kosower,
``Progress in one-loop QCD computations,''
Ann.\ Rev.\ Nucl.\ Part.\ Sci.\  46, 109 (1996)
[hep-ph/9602280].}

\lref\BernFJ{
Z.~Bern, L.~J.~Dixon and D.~A.~Kosower,
``Unitarity-based techniques for one-loop calculations in QCD,''
Nucl.\ Phys.\ Proc.\ Suppl.\  51C, 243 (1996)
[hep-ph/9606378].}

\lref\Catani{S. Catani,
``The Singular Behaviour of QCD Amplitudes at Two-loop Order,''
Phys.Lett. B427 (1998) 161-171 [hep-ph/9802439].}

\lref\parke{S. Parke and T. Taylor, ``An Amplitude For $N$ Gluon
Scattering,'' Phys. Rev. Lett. 56 (1986) 2459}

\lref\giele{F. A. Berends
and W. T. Giele, ``Recursive Calculations For Processes With $N$
Gluons,'' Nucl. Phys. B306 (1988) 759.}








\newsec{Introduction}


Recently there has been renewed interest in the perturbation
expansion of $\N=4$ super Yang-Mills. This was motivated by the
discovery of a twistor string theory \WittenNN\ that captures the
perturbation theory of the maximally supersymmetric Yang-Mills
theory (pMSYM). Twistor string theory has opened new avenues and has
inspired new ideas for the computation of tree level amplitudes of
gluons~\refs{\MHVdec, \treeone, \treetwo, \treethree, \treefour,
\treefive, \treesix, \treeseven, \Higgs, \Higgsone, \Currents,
\RSVNewTree, \recurone, \recurtwo, \LuoWen, \LuoWenone, \BFRSV,
\BadgerMassive} and one-loop amplitudes of gluons in
QCD~\refs{\QCDone, \QCDtwo, \BrandhuberJW},
$\N=1$~\refs{\OtherGaugeCalcs, \oneloopone, \onelooptwo,
\oneloopthree, \oneloopfour, \oneloopfive, \oneloopus} and
$\N=4$~\refs{\BST, \CachazoAnomaly, \BCFseven, \NeqFourSevenPoint,
\General, \NeqFourNMHV}~super Yang-Mills. Before twistor string
theory was introduced, the study of pMSYM at one-loop was mainly
motivated by two facts: one is the decomposition of a QCD amplitude,
$A^{QCD}$, with only a gluon running in the loop in terms of
supersymmetric amplitudes and an amplitude with only a scalar
running in the loop, $A^{\rm scalar}$, (see~\DixonWI\ for a review),
\eqn\sys{A^{\rm QCD}  = A^{\N=4} - 4 A^{\N=1}_{\rm chiral} +
A^{\rm scalar}}
where $A^{\N=4}$ has the full $\N=4$ multiplet in the loop and
$A^{\N=1}_{\rm chiral}$ only an $\N=1$ chiral multiplet. The other
motivation is a surprising proposal of Anastasiou, Bern, Dixon,
and Kosower (ABDK) that two- (and, perhaps, higher-) loop
amplitudes in pMSYM can be completely determined in terms of
one-loop amplitudes \conjecture. This idea was inferred from
studying collinear and IR singular behavior of the higher loop
amplitudes. The conjecture is given in terms of normalized
$2$-loop amplitudes $M^{(2)}_n = A^{(2)}_n/A^{\rm tree}_n$ and in
dimensional regularization, as follows
\eqn\hylo{ M^{(2)}_n(\epsilon) = {1\over 2}\left( M^{(1)
}_n(\epsilon )\right)^2 + f(\epsilon) M^{(1)}_n(2\epsilon ) -{5\over
4}\zeta_4 + {\cal O}(\epsilon).}
This relation was explicitly verified for four-gluon amplitudes in
\conjecture\ (see also section 7 of~\hundred). Also based on
collinear limits~\Kosower, the schematic form of a relation
analogous to~\hylo\ was proposed for higher loops \high. Very
recently, an explicit formula, analogous to \hylo, for the
three-loop four-gluon amplitude was obtained and successfully
verified in~\New. It is the aim of this paper to make some modest
steps towards the calculation of higher loop amplitudes in pMSYM.
The main motivation is to prepare the ground for future tests of
the ABDK proposal. A proof of \hylo\ would lead to the solution of
pMSYM at two loops as a general solution to the one-loop problem
can be obtained in terms of tree-level amplitudes by using
quadruple cuts~\General. This is possible thanks to the cut
constructibility of one-loop amplitudes in pMSYM proven in \BernZX\
and the decomposition in terms of scalar box integrals, with
rational functions as coefficients, also given in \BernZX. See
also ~\refs{\BST, \CachazoAnomaly, \BCFseven, \NeqFourSevenPoint,
\NeqFourNMHV} for other techniques in pMSYM at one loop.

At two loops, a similar decomposition in terms of some given set
of integrals is expected by using Passarino-Veltman or similar
reduction procedures~\passarino. Unfortunately, the complete basis
of two-loop integrals is currently unknown. However, scalar double
box integrals are a natural ingredient of such a basis\foot{In
fact, the four-gluon amplitude is given only in terms of scalar
double boxes \Bernry.}. In this paper, we concentrate on the
calculation of the coefficient of certain classes of planar scalar
double box integrals. These are the integrals that arise in scalar
field theory with a massless scalar running along internal lines
and with the double-box structure depicted in fig. 1.
\ifig\topo{The two possible different structures of planar scalar
double box integrals. $(a)$ Double boxes. $(b)$ Split double
boxes. Note that the momenta of the external lines is given by the
sum of the momenta of external gluons.}
{\epsfxsize=0.65\hsize\epsfbox{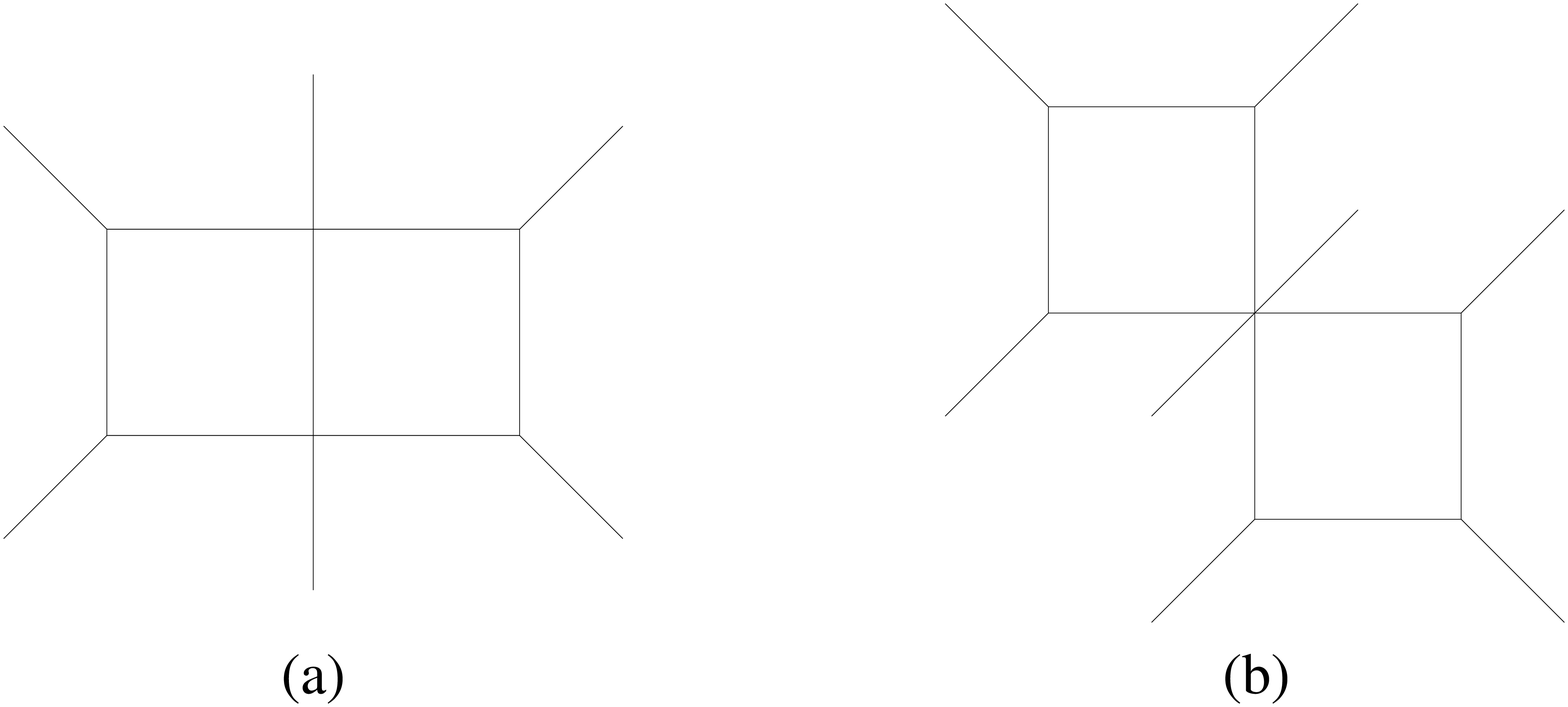}}
\noindent
The momenta of the external legs in fig. 1 are given by sums of
momenta of external gluons.

We propose a method for computing the coefficient of any scalar
double box integral given in fig. 1a when at least one of the two
boxes has two adjacent massless three-particle vertices. We also
give the form of the coefficient of any double box given in fig.
1b. In order to distinguish between the double boxes in fig. 1a
and in fig. 1b we refer to the former simply as ``double boxes"
and the latter as ``split double boxes".

Our original motivation was the successful use of quadruple cuts
in the calculation of one-loop $\N=4$ amplitudes~\General. The
basic idea is that at one-loop only scalar boxes contribute
\BernZX. A quadruple cut singles out the contribution of a given
scalar box and localizes the integration over the loop momentum.
The combination of these two facts allows one to calculate any
coefficient in terms of the product of four tree-level amplitudes
\General. Up to a numerical factor, every one-loop box coefficient
is given by
\eqn\cuad{  B = \sum A_{(1)}^{\rm tree}A_{(2)}^{\rm
tree} A_{(3)}^{\rm tree}A_{(4)}^{\rm tree},}
where the sum is over the solutions to the delta function
equations and over all particles that can propagate in the loop. A
straightforward application of this idea can be made for split
double boxes (see fig. 1b). Again the idea is to cut all eight
propagators, i.e., an octa-cut which localizes the two loop
integrations and gives the coefficient as the product of seven
tree-level amplitudes (up to a numerical factor)
\eqn\douge{ B = \sum \prod_{i=1}^7 A_{(i)}^{\rm tree}.}
Naively, one might expect that the coefficient of double boxes in
fig. 1a cannot be computed in a similar manner. The reason is that
there are only seven propagators and a hepta-cut does not localize
the integrals over the loop momenta.

A way to avoid the remaining integration arises in an unexpected
manner. In studying singularities of Feynman integrals, one computes
the discontinuity of an integral across a singularity by cutting
propagators. When one cuts all propagators in a Feynman diagram one
is computing the discontinuity across the leading singularity of the
integral. However, at two (and higher) loops one finds a surprise
when some of the external legs are massless. At two loops, if any of
the two boxes in fig. 1a has at least two adjacent three-particle
vertices (condition that is satisfied trivially for less than seven
external gluons), then the integral has an extra propagator-like
singularity beyond the naive leading singularity. The discontinuity
across the new leading singularity is actually computed by an
octa-cut\foot{For more general double boxes, there is also an extra
singularity, these are known as second-type singularities~\thebook.
They cannot easily be used to produce an octa-cut but they might
give a generalization of it.}. This octa-cut precisely localizes the
loop integrations and allows a straightforward computation of the
coefficient as the product of six tree-level amplitudes. Up to a
numerical factor, it is given by
\eqn\sixtree{ B = \sum \prod_{i=1}^6 A_{(i)}^{\rm
tree}.}

The only two-loop amplitude in pMSYM known in the literature is
the four-gluon amplitude \Bernry. One reason is that very few
double scalar box integrals are known explicitly \smirnov. In
particular, to our knowledge, not all double box integrals needed
for a five-gluon amplitude are known. Nevertheless, we present the
computation of several five-gluon and six-gluon non-MHV scalar
double box integrals as illustrations of our technique.

This paper is organized as follows. In section 2, we review pMSYM
at tree-, one-, two- and three-loop levels as well as the ABDK
conjecture. In section 3, we show that the four-gluon amplitude of
pMSYM can be found by using hepta-cuts. Even though the number of
cut propagators is less than the number of integration variables,
the integrand turns out not to depend on the loop momenta and can
be pulled out of the integral. In section 4, we demonstrate that a
certain class of double-box configurations admit an extra
propagator-type singularity. Cutting this singularity allows us to
write a universal formula for many double-box coefficients. In
section 5, we illustrate our technique via various examples
including non-MHV amplitudes. In section 6, we discuss
applications to three- and higher-loop amplitudes. In particular,
we show that by studying singularities, it is possible to realize
that the basis of integrals has to contain integrals with some
non-trivial factors in the numerator. This is in agreement with
results of~\Bernry.

Throughout the paper, we use the following notation and
conventions along with those of \WittenNN\ and the spinor
helicity-formalism \refs{\berends,\xu,\gunion}. A external gluon
labeled by $i$ carries momentum $K_i$. Since $K_i^2=0$, it can be
written as a bispinor $(K_i)_{a\dot a} =
\lambda_{i~a}\tilde\lambda_{i~\dot a}$. Inner product of null
vectors $p_{a\dot a}=\lambda_a \lt_{\dot a}$ and $q_{a\dot
a}=\l'_a \lt'_{\dot a}$ can be written as $2p\cdot q = \vev{\l ,
\l'}[\lt ,\lt' ]$, where $\vev{\l , \l' } =
\epsilon_{ab}\l^a\l'^b$ and $[\lt , \lt' ] = \epsilon_{\dot a\dot
b}\lt^{\dot a}\lt'^{\dot b}$. Other useful definitions are:
\eqn\ournot{ \eqalign{ K_{i,\ldots ,j} &\equiv
K_i+K_{i+1}+\cdots+K_{j}\ \cr K_i^{[r]} &\equiv
K_i+K_{i+1}+\cdots+K_{i+r-1}\ \cr \gb{i|\sum_r  K_r |j} &\equiv
\sum_r \vev{i~r}[r~j] \cr \vev{i|(\sum_r K_r )(\sum_s K_s )|j}
&\equiv \sum_r\sum_s \vev{i~r}[r~s]\vev{s~j} \cr [i|(\sum_r K_r
)(\sum_s K_s )|j] &\equiv \sum_r\sum_s [i~r]\vev{r~s}[s~j] \cr
\gb{i|(\sum_r K_r)(\sum_s K_s )(\sum_t K_t )|j} &\equiv \sum_r
\sum_s \sum_t \vev{i~r}[r~s]\vev{s~t}[t~j] }}
where addition of indices is always done modulo $n$.


\newsec{Review of $\N=4$ Amplitudes}


In this paper we consider amplitudes of gluons in ${\cal N}=4$
super-Yang-Mills. Each gluon carries the following information:
momentum $p_{a\dot a}$, polarization vector $\epsilon_{a\dot a}$
and color index $a$. The color structure can be striped out by a
color decomposition
\refs{\colorordone,\colorordtwo,\colorordthree,\colorordfour}.
Here we only consider the leading color or planar part of the
amplitudes. The information in momentum and polarization vectors
can be encoded in terms of spinors $\lambda$, $\tilde\lambda$ and
the helicity of the gluon $h$.


\subsec{Tree-Level $\N=4$ Amplitudes}


At tree-level, the leading color approximation is exact. An
amplitude is given by
\eqn\caso{A_{(\{ p_i,\epsilon_i,a_i \})} = g_{\rm
YM}^{n-2}\sum_{\sigma\in S_n / Z_n} \Tr (T^{a_{\sigma(1)}}\ldots
T^{a_{\sigma(n)}})A_{( \{ \lambda_{\sigma(1)},
\tilde\lambda_{\sigma(1)} , h_{\sigma(1)} \} ,\ldots , \{
\lambda_{\sigma(n)}, \tilde\lambda_{\sigma(n)} , h_{\sigma(n)}
\})}. }
Here we are suppressing a delta function that imposes momentum
conservation.

It is convenient to denote the set of data $\{ \lambda_i,
\tilde\lambda_i , h_i \}$ by $i^{h_i}$, where $h_i = \pm$ is the
helicity of the $i^{th}$ gluon. The amplitudes on the right hand
side of \caso\ are known as leading color partial amplitudes and
are computed from color-ordered Feynman rules. One can study a
given order $A(1^{h_1},\ldots , n^{h_n})$ and the rest can be
obtained by application of permutations, $\sigma$.

The partial amplitude $A(1^{h_1},\ldots , n^{h_n})$ can be computed
using a variety of methods (see \DixonWI\ for a nice review on many
of the techniques developed in the 80's and 90's). More recently,
two new techniques became available, namely, MHV diagrams \MHVdec\
and the BCFW recursion relations \refs{\recurone,\recurtwo}. The
latter is a set of quadratic recursion relations for on-shell
physical partial amplitudes of gluons. For a recent review see
\cape.


\subsec{One-Loop $\N=4$ Amplitudes}


Amplitudes of gluons at one-loop admit a color decomposition
\refs{\colorordone,\colorordtwo,\colorordthree,\colorordfour,\mangparke}
with single and double trace contributions. As mentioned in the
introduction we will only concentrate on the leading color partial
amplitudes\foot{It is interesting to note that since for $\N=4$
SYM all particles in the loop are in the adjoint representation,
all sub-leading color amplitudes are given as linear combinations
of the planar ones with permutations of the gluon labels (See
section 7 of \BernZX\ for a proof.)}.

One-loop amplitudes of gluons in supersymmetric theories are
four-dimensional cut-constructible~\refs{\BernZX,\Fus}. This means
that they can be completely determined by their finite branch cuts
and discontinuities.
$\N=4$ amplitudes are even more special. Reduction
techniques~\passarino\ can be used to express these amplitudes in
terms of scalar box integrals \BernZX. These are one-loop box
Feynman integrals in a scalar field theory where a massless scalar
runs in the loop,
\eqn\ifout{{\cal I} = \int d^4\ell { 1\over (\ell^2+i\epsilon)
((\ell-k_1)^2 + i\epsilon )((\ell -k_1-k_2)^2 + i\epsilon
)((\ell+k_4)^2 + i\epsilon )}}
where $k_1,k_2,k_3,k_4$ are the external momenta at each vertex.
They are not independent since by momentum conservation $k_3 =
-(k_4+k_1+k_2)$. Note that the integral \ifout\ is singular when
at least one $k_i$ is a null vector. Therefore, we should specify
a regularization procedure, like dimensional regularization.
However, we will be considering cuts that are finite and do not
depend on the regularization procedure. Since $A(1,\ldots, n)$ is
color-ordered, each $k$ can only be the sum of consecutive momenta
of external gluons. Moreover, since we only consider the planar
contributions, we can define a given contribution by specifying
$i,j,k,l$ such that $k_1 = K_i+\ldots + K_{j-1}$, $k_2= K_j+\ldots
+ K_{k-1}$ and $k_3 = K_k+\ldots + K_{l-1}$. The reduction
procedure then gives for the amplitude an expansion of the form
\BernZX\
\eqn\exas{A(1,\ldots ,n ) = \sum_{1<i<j<k<m<n} B_{ijkl} ~ {\cal
I}_{(K_{i}+\ldots + K_{j-1}, K_{j}+\ldots +K_{k-1}, K_{k}+\ldots
+K_{l-1})}, }
where the coefficients $B_{ijkm}$ are rational functions of the
spinor products.
Since all scalar box integrals are known explicitly, the problem
of computing $A(1,\ldots ,n)$ is reduced to that of computing the
coefficients $B_{ijkl}$.

A general formula for the coefficients $B_{ijkl}$ was found in
\General\ in terms of products of tree level amplitudes. Let us
review the derivation of the formula because the idea is useful in
the analysis at higher loops. If we think of the scalar box
integrals as an independent basis\foot{The notion of independence
is the equivalent of cut constructibility of the amplitude.} of
some vector space we can interpret $A(1,\ldots ,n)$ as a general
vector. All we need to do is to find a way to project $A(1,\ldots
,n)$ along the space spanned by a given scalar box integral ${\cal
I}$. {}From the definition of ${\cal I}$ in \ifout\ we see that
each integral is uniquely determined once its four propagators are
given. It is natural to think that the way to determine the
coefficient $B$ is by looking at the region of integration where
all four propagators become singular. In fact, the integral
obtained by cutting, i.e., by dropping the principal part of all
four propagators computes the discontinuity of the given scalar
box integral across a singularity which is unique to it.

The set of four equations that gives $\ell$ is the following
\eqn\erq{\ell^2=0, \qquad (\ell-k_1)^2=0 , \qquad
(\ell-k_1-k_2)^2=0, \qquad (\ell+k_4)^2 = 0.}
A little exercise shows that these equations do not have a
solution if $\ell$ is a real vector in Minkowski space for general
external momenta. The way out of this problem is to complexify all
momenta and make a Wick rotation to $(- - + +)$ signature. In the
new signature the delta functions are still well defined and there
are always solutions to \erq.

\ifig\druple{A quadruple cut diagram. Momenta in the cut
propagators flows clockwise and external momenta are taken
outgoing.  The tree-level amplitude $A^{\rm tree}_1$, for example,
has external momenta $\{ i+1,...,j,\ell_2,\ell_1 \}$.}
{\epsfxsize=0.40\hsize\epsfbox{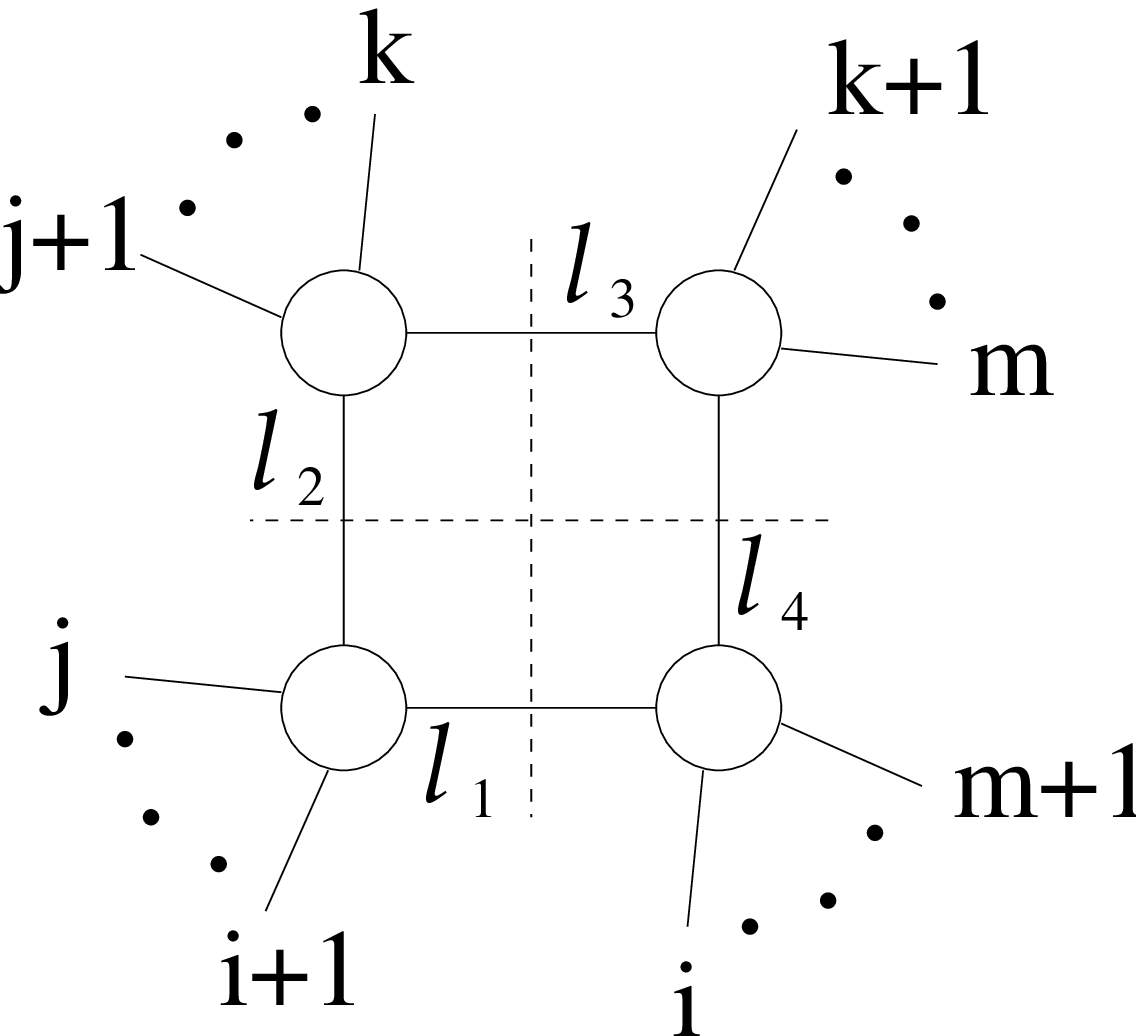}}
\noindent One can also look at the same regime on the left hand
side of \exas\ by considering only Feynman diagrams that posses
the four propagators entering in \erq. Summing over them one finds
the following equation
\eqn\cuti{ \int d\mu \sum_{J} n_J A^{tree}_{(1)} A^{tree}_{(2)}
A^{tree}_{(3)} A^{tree}_{(4)} = B_{ijkm} \int d\mu  }
where sum over $J$ represents a sum over all possible particles in
the $\N=4$ multiplet. The measure $d\mu$ is the same one both
sides of the integrals,
\eqn\qeqi{d\mu = d^4\ell ~\delta^{(+)}(\ell^2)
~\delta^{(+)}((\ell-k_1)^2) ~\delta^{(+)}((\ell
-k_1-k_2)^2)~\delta^{(+)}((\ell+k_4)^2),}
and the tree-level amplitudes are defined as follows (see~\druple)
\eqn\treeam{\eqalign{ A_{(1)}^{tree} = & A(-\ell_1,i+1,i+2,\ldots
,j-1,j,\ell_2 ),  \quad A_{(2)}^{tree} =  A(-\ell_2 ,
j+1,j+2,\ldots ,k-1, k, \ell_3), \cr A_{(3)}^{tree} = & A(-\ell_3,
k+1,k+2,\ldots , m-1,m,\ell_4),  ~ A_{(4)}^{tree} = A(-\ell_4,
m+1,m+2,\ldots ,i-1,i, \ell_1).} }
where
\eqn\deff{\eqalign{& \ell_1 = \ell, ~ \ell_2 = \ell -k_1, ~\ell_3
= \ell - k_1 - k_2, ~ \ell_4 = \ell + k_4, ~ k_1 = K_{i+1}+\ldots
+ K_{j}, \cr &  k_2 = K_{j+1}+\ldots + K_{k}, ~ k_3 =
K_{k+1}+\ldots + K_m, ~ k_4=K_{m+1}+\ldots + K_i. }}
The integral $\int d\mu$ is just given by a Jacobian
$1/\sqrt{\Delta}$. This Jacobian cancels on both sides since the
integral is localized by the delta functions and the coefficient
is given by \General\
\eqn\cool{ B_{ijkl} = {1\over |{\cal S}|}\sum_{{\cal S},J} n_J
A^{tree}_{(1)} A^{tree}_{(2)} A^{tree}_{(3)} A^{tree}_{(4)}. }
Here ${\cal S}$ is the set of solutions to the conditions imposed
by the delta functions, and $|{\cal S} |$ is the number of
solutions. The sum also involves a sum over all possible particles
that can propagate in the loop. For further details and many
examples we refer to~\General. Even though the Jacobian did not
play an important role for the quadruple cut technique at
one-loop, it is crucial for the two-loop analysis we carry out in
section 4.1. For this reason let us write it down for future
reference
\eqn\gram{\Delta = s^2t^2-2 s t
(k_1^2k_3^2+k_2^2k_4^2)+(k_1^2k_3^2-k_2^2k_4^2)^2}
with $s= (k_1+k_2)^2$ and $t=(k_2+k_3)^2$.


\subsec{Two-Loop $\N=4$ Amplitudes}


At two loops, only the four-gluon amplitude has been
computed~\Bernry. The $\N=4$ calculation was the first full
two-loop amplitude of gluons ever computed. The answer is given as
a linear combination of double box scalar integrals with
coefficients that are rational function of the spinor variables. A
double box scalar integral is the analog of the one-loop scalar
box integral introduced above, more explicitly,
\eqn\eeone{\eqalign{ {\cal I}(k_1,\ldots , k_6) = & \int {d^4p
\over (2\pi)^4} {1 \over
(p^2+i\epsilon)((p-k_1)^2+i\epsilon)((p-k_1-k_2)^2+i\epsilon)}\times
\cr & \int {d^4 q\over (2\pi)^4} {1\over
((p+q+k_6)^2+i\epsilon)(q^2+i\epsilon)((q-k_5)^2+i\epsilon)((q-k_4-k_5)^2+i\epsilon)}.}}
This integral is UV finite but it might have IR divergences when
some $k$'s are null vectors. Again, as in the one-loop case, one
has to choose a regularization procedure but we do not do so
because we only discuss finite cuts. The planar contribution to
the four-gluon amplitude is~\Bernry\
\eqn\fourtwo{ A^{2-loop}_4(K_1,K_2,K_3,K_4) = A^{tree}_4 ~s~t
\left( s ~ {\cal I}(K_1,K_2,0,K_3,K_4,0) + t ~ {\cal
I}(K_4,K_1,0,K_2,K_3,0) \right) }
where $s=(K_1+K_2)^2$ and $t=(K_2+K_3)^2$. This was computed by
using the unitarity-based method \refs{\BernZX, \Fus, \BernDB,
\BernFJ, \BernJE}. It is very important to mention that the double
box scalar integral \eeone\ is not known in general but explicitly
formulas exists in dimensional regularization when $k_3=k_6=0$ and
at least three of the other $k_i$'s are null vectors ~\smirnov.


\subsec{ABDK Conjecture}


As mentioned in the introduction, one of the motivations of this
work is to prepare the ground for a more extensive test of the
ABDK conjecture. This conjecture asserts that the planar limit of
$L$-loop amplitudes in $\N=4$ SYM is determined iteratively, i.e.,
as a function of $l$-loop amplitudes with $l<L$.

Let us make this more precise. Here we follow~\conjecture\
and~\high\ where the original proposal was made. Consider the
function
\eqn\ratio{ M^{(L)}_n (1,2,\ldots ,n)= {A^{L-loop}(1,2,\ldots ,n )
\over A^{tree}(1,2,\ldots , n)}}
then the ABDK conjecture states that
\eqn\rety{ M^{(L)}_n = P_L(M_n^{(1)}, \ldots , M_n^{(L-1)})}
where $P_L(x_1,\ldots ,x_{L-1})$ is a certain polynomial of degree
$L$ and independent of the helicity configuration. The explicit
form of \rety\ at two loops was given in~\conjecture\ in terms of
the function $f(\epsilon ) = (\psi ( 1 - \epsilon )-\psi
(1))/\epsilon$, where the digamma function is defined by $\psi(x)
= \Gamma'(x)/\Gamma(x)$, as follows
\eqn\abdk{ M_n^{(2)}(\epsilon ) = {1\over 2}\left( M_n^{(1)}(\epsilon)
\right)^2 + f(\epsilon )M_n^{(1)}(2\epsilon ) - {5\over 4}\zeta_4
.}
This conjecture was explicitly checked for four-gluon amplitudes.
Very recently, the form of the polynomial in~\rety\ was obtained
for the three-loop four-gluon amplitude in~\New. One of the
impressive predictions of the conjecture is a relation between the
finite remainders which are defined at $\epsilon =0$. At two
loops, one introduces the universal singular function
$C_n(\epsilon)^{(2)}$~\refs{\Catani, \conjecture} which contains
the infrared singularities and does not depend on the helicity
configuration since it is normalized by the tree-level amplitude.
Defining the finite remainder as
\eqn\rema{F_n^{(2)}(\epsilon ) = M^{(2)}_n(\epsilon ) -
C^{(2)}_n(\epsilon ),}
one can write a finite (as $\epsilon \to 0$ ) analog of~\abdk\ as follows~\conjecture
\eqn\conj{ F_n^{(2)}(0) = {1\over 2} \left( F_n^{(1)}(0)\right)^2
 - \zeta_2 F_n^{(1)}(0) - {1\over 4}\left( {11n\over 8} + 5 \right) \zeta_4.}
Recall that at one-loop $F_n^{(1)}(0)$ can have at most
dilogarithms, while $F_n^{(2)}(0)$ can have higher polylogarithms.
This means that very non-trivial cancelations must happen. These
cancelations were found to occur for $n=4$ between terms coming from
the two integrals in \fourtwo\ and involved many polylogarithmic
identities~\conjecture. In the recent paper \New, an impressive
formula for the all loop finite remainder of MHV amplitudes was also
presented. The formula is given in a kind of generating function
structure
\eqn\wowe{1+\sum_{L=1}^{\infty} a^L F_n^{(L)}(0) = {\rm exp}\left[
{1\over 4}\gamma_K F_n^{(1)}(0) + C \right] }
where $a$ is basically the 't Hooft coupling, $\gamma_K$ is the
universal soft anomalous dimension and $C$ is a function that admits
a power series representation in $a$. $\gamma_K$ and $C$ are known
up to the order needed to obtain the three-loop term\foot{It is
important to mention that there is no canonical definition of the
finite remainder $F_n$. In fact, the definition of finite remainder
used in \conj\ differs from that used in \wowe. For more details see
\New. We thank Z. Bern and L. Dixon for useful discussions on this point.}.


\newsec{Four-Gluon Two-Loop Amplitudes and Hepta-Cuts}


In this section, we consider hepta-cuts of the two-loop four-gluon
leading partial amplitude. This section can be viewed as a warm-up
section where we introduce relevant notation and do some sample
calculations which will be used in the rest of the paper. It is
enough to consider $A_4^{2-loop}(1^-, 2^-, 3^+, 4^+)$ as all other
$A_4^{2-loop}$ with different helicity assignments can be obtained
from this one by Ward identities. The leading partial amplitude
was first computed in~\Bernry. As reviewed in section 2.3, the
amplitude can be presented as a linear combination of two scalar
double-box integrals (see fig. 1a)
\eqn\eone{{\cal I} =\int {d^{4}p\over (2\pi)^4}{d^{4}q\over
(2\pi)^4} {1 \over
p^2(p-K_1)^2(p-K_1-K_2)^2(p+q)^2q^2(q-K_4)^2(q-K_3-K_4)^2},}
where $K_i$ are the four external gluon momenta, with rational
coefficients. All external momenta are assumed to be outgoing. The
integral~\eone\ has seven propagators, hence it is natural to
consider hepta-cuts. It turns out that the coefficients can easily
be found from hepta-cuts when the loop momenta are analytically
continued to signature $(- - + +)$ or complexified. In the present
case, there are two independent coefficients as well as two
independent hepta-cuts. We refer to them as the $s$-channel cut
and the $t$-channel cut. The corresponding coefficients will be
denoted as $c_s$ and $c_t$. Let us start with the cut in the
$s$-channel. In this case, there are six different helicity
configurations. For all of them, only gluons can propagate in both
loops. A sample helicity configuration is shown in fig. 3. In this
paper, for simplicity, we depict tree level amplitudes as points.
Since all propagator are cut, there is no need to indicate a cut
by a dash line and we choose to omit
them~\foot{Note that conventions in fig. 3 are different from those used
in fig. 2 where all tree level
amplitudes are denoted by blobs and cuts are indicated by dashed
lines.}.
\ifig\figureone{A sample hepta-cut in the $s$-channel. Tree level
amplitudes are depicted as points. All propagators are cut and
therefore we omit the dashed lines used in fig. 2.}
{\epsfxsize=0.50\hsize\epsfbox{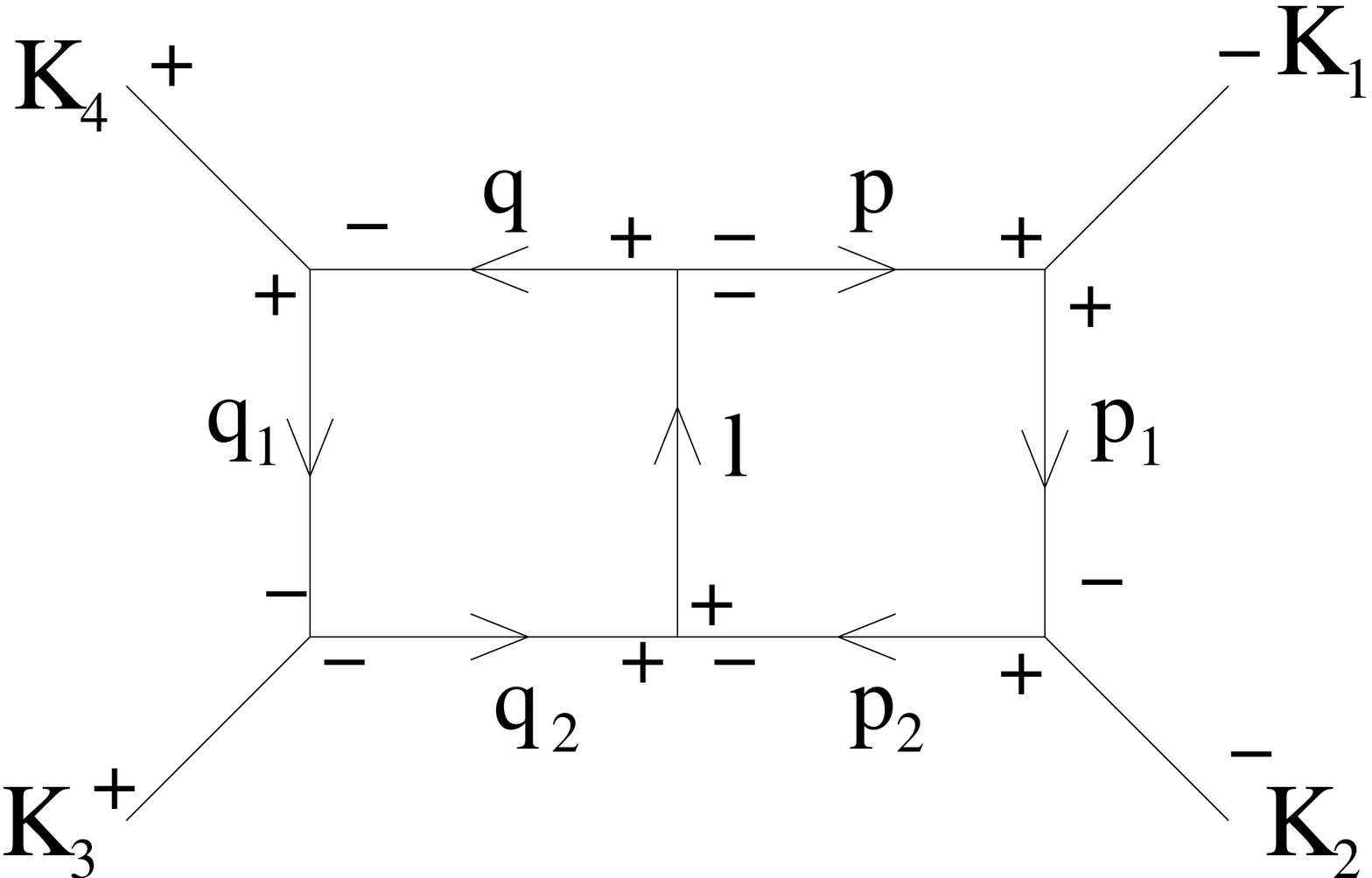}}
\noindent
The rational coefficient $c_s$ is then given by
\eqn\etwo{
c_s = {i^7\sum_{I=1}^6 \int d\mu
(A_{(1)}^{tree}A_{(2)}^{tree}A_{(3)}^{tree}A_{(4)}^{tree}A_{(5)}^{tree}A_{(6)}^{tree})_I
\over \int d\mu},}
where by $A_{(i)}^{tree}$ we denote tree amplitudes at each of the six vertices,
the integration measure $d\mu$ is given by
\eqn\ethree{\eqalign{ d\mu =& {d^4p \over (2\pi)^4} {d^4 q \over
(2\pi)^4} \delta(p^2) \delta ((p-K_1)^2) \delta ((p-K_1-K_2)^2)
\delta ((p+q)^2) \cr & \delta (q^2) \delta((q-K_4)^2) \delta
((q-K_3-K_4)^2)}}
and in the numerator we sum over the product of the six tree level
amplitudes corresponding to a given helicity configuration. The
factor $i^7$ comes from the seven propagators. It seems that since
the number of delta functions is less than the number of
integration variables, the integral does not localize and one
integration has to be performed. However, it turns out that the
integrand can be simplified in such a way that the dependence on
the loop momenta drops out and we are left with the integral of
the measure which cancels out according to eq.~\ethree.

In the discussion of one-loop amplitudes in section 2.2, we
mentioned that the momentum $\ell$ has to be complexified in order
to find solutions to the four equations from the cut propagators.
Making $\ell$ complex also has as a byproduct the fact that
three-particle vertices on-shell do not have to vanish. Tree-level
three-gluon amplitudes with helicities $(- - +)$ or $(+ + -)$ are
given respectively by~\refs{\parke,\giele}
\eqn\efour{
A_3^{tree}(p^-, q^-, r^+)=
i{\vev{p~q}^3 \over \vev{q~r} \vev{r~p}}, \quad
A_3^{tree}(p^+, q^+, r^-)=-i
{[p~q]^3 \over [q~r] [r~p]}.}
In Minkowski space, $\lambda_p$ and $\tilde{\lambda}_p$ are
related to each other as $\tilde{\lambda}_p=\pm \bar \lambda_p$.
This means that if $p\cdot q=0$, which follows from momentum
conservation at the vertex, then both
$\vev{\lambda_p~\lambda_{q}}=0$ and
$[\tilde{\lambda}_{p}~\tilde{\lambda}_{q}]=0$. This implies that
both amplitudes in~\efour\ vanish. If we complexify the momenta,
then the equation $p\cdot q=0$ has two independent solutions. We
have that either $\vev{\lambda_p~\lambda_{q}}=0$ or
$[\tilde{\lambda}_{p}~\tilde{\lambda}_{q}]=0$. That is either
${\lambda_p}$ and $\lambda_{q}$ are proportional or
$\tilde{\lambda}_{p}$ and $\tilde{\lambda}_{q}$ are proportional.
Also note that momentum conservation implies that $p\cdot q=p\cdot
r =q\cdot r =0$. This means that either three $\lambda$'s are
proportional or all three $\tilde{\lambda}$'s are proportional.
Therefore, for every $(+ + -)$ tree level amplitude we choose all
$\lambda$'s to be proportional. Similarly, for every $(- - +)$
tree level amplitude we choose all three $\tilde{\lambda}$'s to be
proportional.

Explicit calculations, considered for one of the helicity
configurations in some detail below, show that every helicity
configuration gives the same contribution equal to
\eqn\efive{
-A_4^{tree}s^2t \int d\mu,}
where $A_4^{tree}$ is the tree-level four-gluon amplitude
\eqn\esix{
A_4^{tree}(1^-, 2^-, 3^+, 4^+)=i{\vev{1~2}^3 \over \vev{2~3}\vev{3~4}\vev{4~1}},}
and
\eqn\eseven{ s=(K_1+K_2)^2, ~ t=(K_2+K_3)^2.}

Note that the integral in \efive\ cancels against the denominator
in~\etwo. The coefficient $6$ in the numerator will also cancel.
The reason is the following. In the denominator in~\etwo, we have
to sum over all different solutions to the delta-function
conditions. It is easy to realize that in this particular case the
number of different solutions equals the number of helicity
configurations. Thus, each term in the numerator in~\etwo\ picks
one of the six solutions whereas in the denominator we sum over
all the six solutions. As a result, we obtain
\eqn\enine{
c_s=-A_4^{tree}s^2t.}
This coincides with the corresponding coefficient found in~\Bernry.

Let us consider the helicity configuration shown in fig. 3 in some
detail. The analysis of the remaining five configurations is
completely analogous. Consider the integrand as the product of six
tree-level amplitudes
\eqn\eten{\eqalign{ &
-i(A_{(1)}^{tree}A_{(2)}^{tree}A_{(3)}^{tree}A_{(4)}^{tree}A_{(5)}^{tree}A_{(6)}^{tree})_1=
-i \left( {[p_1, p]^3 \over [p, 1] [1, p_1]} \right)
\left({\vev{p_1, 2}^3 \over \vev{2, p_2} \vev{p_2, p_1}}\right)
\cr & \left( {[q_2, l]^3 \over [l, p_2] [p_2, q_2]} \right)
\left({\vev{p, l}^3 \over \vev{l, q} \vev{q, p}}\right) \left(
{[q_1, 4]^3 \over [4, q] [q, q_1]} \right) \left({\vev{q_1, q_2}^3
\over \vev{q_2, 3} \vev{3, q1}}\right).}}
Next, simplify this expression by using momentum conservation. For
example, the product of $[p_1, p]$ and $\vev{p_1, 2}$ can be
simplified as follows
\eqn\eeleven{
[p_1, p] \vev{p_1, 2} =-\gb{2|p_1|p}=\gb{2|K_1|p}=-\vev{1~2}[1~p].}
Then the product of the first four factors in~\eten\ becomes
\eqn\etwelve{ \vev{1~2}^2[q, q_2]^2.}
After using momentum conservation along the lines of eq.~\eeleven,
one finds
\eqn\ethirteen{
-i(A_{(1)}^{tree}A_{(2)}^{tree}A_{(3)}^{tree}A_{(4)}^{tree}A_{(5)}^{tree}A_{(6)}^{tree})_1=
i s^2 {\vev{1~2}^2[3~4] \over \vev{3~4}} =-A_4^{tree}s^2 t.}
Note that this expression does not depend on the loop momenta and, thus, can be
pulled out of the integration.

Now we consider the hepta-cut in the $t$-channel. Here we have ten
helicity configurations. Note that in this case the number of
helicity configurations does not equal the number of solutions of
the delta-function equations. By a solution we mean a choice
whether all $\lambda$'s are proportional or $\tilde{\lambda}$'s
are proportional at each of six three gluon vertices. However,
among the ten configurations, there are different configurations
for which the choices of whether $\lambda$'s or
$\tilde{\lambda}$'s are proportional are exactly the same. A
solution then means summing up over such configurations. In this
case, there are two helicity configurations corresponding to
actual solutions and the remaining eight ones break up in pairs.
The sum of the two helicity configurations in each pair
corresponds to a solution of the delta-function conditions.
Overall, we have six improved helicity configurations, each
corresponding to an independent solution to the delta-function
conditions. All paired up configurations involve fermions and
scalars running in one of the loops and summation over the two
configurations in a given pair provides a significant
simplification. The coefficient $c_t$ is given by
\eqn\efourteen{ c_t = -{i\sum_{I=1}^{10} \int d\mu
(A_{(1)}^{tree}A_{(2)}^{tree}A_{(3)}^{tree}A_{(4)}^{tree}A_{(5)}^{tree}A_{(6)}^{tree})_I
\over \int d\mu},}
where the sum is over all ten configurations, or over the six
improved configurations, each corresponding to an actual solution
of the delta-function equations. As before, all six improved
configurations give the same contribution
\eqn\efifteen{
-A_{4}^{tree}st^2.}
Since each improved configuration corresponds to a solution to the
delta-function equations, the factor $6$ cancels out. As a result
we obtain
\eqn\esixteen{
c_{t}=-A_{4}^{tree}st^2.}
This coincides with the corresponding coefficient from~\Bernry. As
an example, let us consider the two helicity configurations shown
in fig. 4.
\ifig\figuretwo{Examples of hepta-cuts in the $t$-channel that
correspond to the same solution of the delta function
constraints.}{\epsfxsize=0.60\hsize\epsfbox{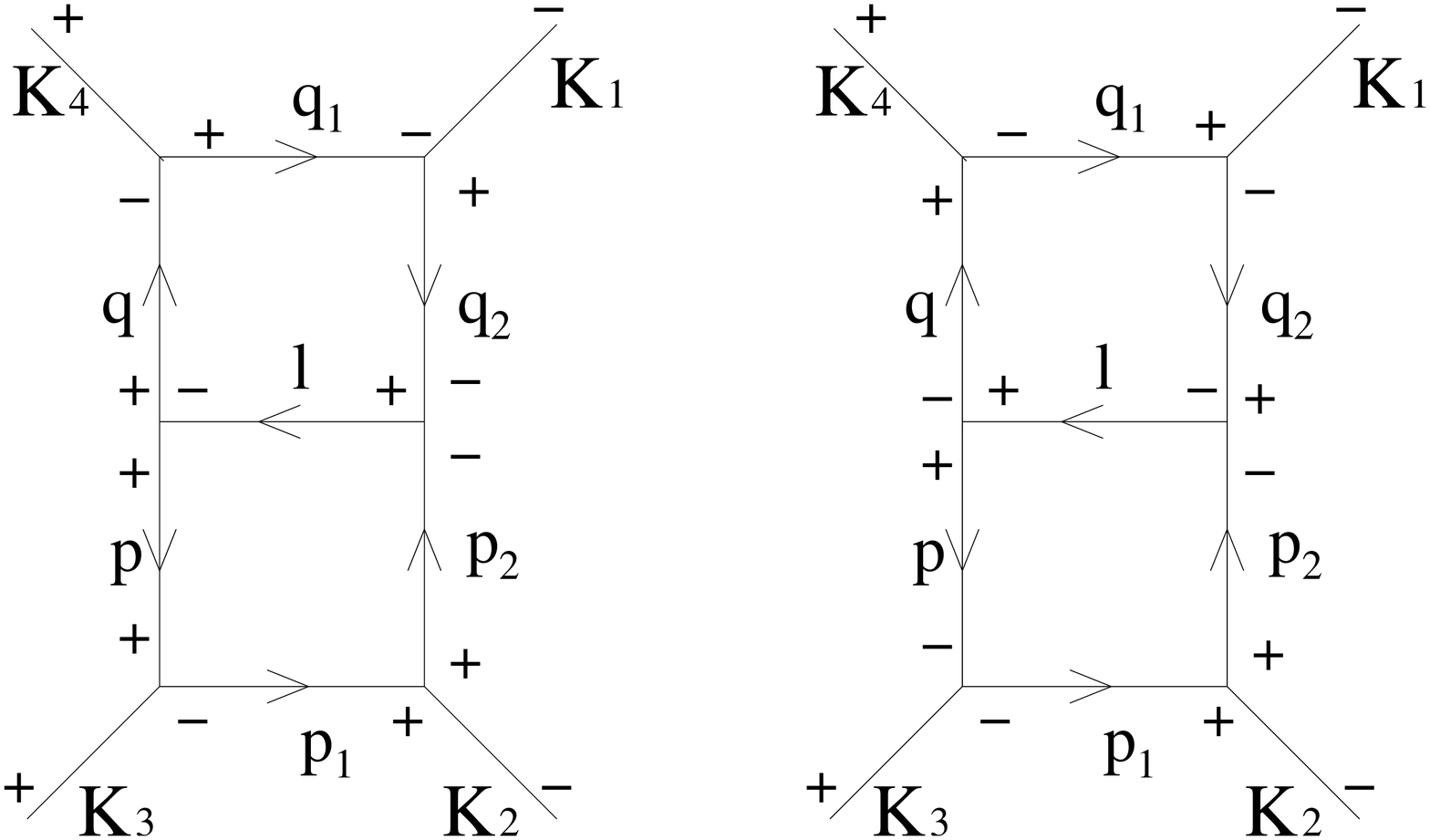}}
\noindent
Note that for both double boxes the choices
whether all $\lambda$'s or all $\tilde{\lambda}$'s are proportional
are exactly the same at every vertex. Therefore,
it is the sum of these two diagrams
that corresponds to one of the six improved helicity configurations.
Both double boxes in fig. 4 involve gluons, fermions and scalars
running in one of the loops and only gluons running in the remaining loop.
The necessary tree level amplitudes are given by
\eqn\eseventeen{
A_3^{tree}(p^-, q^-, r^+)=
i{\vev{p~q}^3 \over \vev{q~r} \vev{r~p}} \left({\vev{q~r} \over \vev{q~p}}\right)^a, \quad
A_3^{tree}(p^+, q^+, r^-)=-i
{[p~q]^3 \over [q~r] [r~p]}\left( {[q~r]\over [q~p]}\right)^a,}
where $a=0$ for gluons, $a=1$ for fermions and $a=2$ for scalars.
After summation over the two configurations, we obtain
\eqn\eeighteen{
-i{(\alpha-\beta)^4 \over \gamma},}
where
\eqn\eninteen{\eqalign{
& \alpha=\vev{q_1~1}\vev{q_2~p_2}[p~q][4~q_1], \quad
\beta=\vev{1~q_2}\vev{p_2~l}[l~p][q~4] \cr &
\gamma=\vev{q_2~q_1}\vev{l~q_2}[q~l][q_1~q] \vev{1~q_2}\vev{p_2~l}[l~p][q~4]
\vev{q~1} \vev{q_2 p_2}[4~q_1][p~q].}}
By using momentum conservation along the lines of eq.~\eeleven, we
can simplify~\eninteen\ to obtain $-A_{tree}st^2$. The integral of
the measure factors out and cancels against the denominator
in~\efourteen\ to give \esixteen.

Thus, we find that the coefficients of the four-gluon amplitude
double boxes can be calculated by studying hepta-cuts. Of course,
this is not enough to claim that this is the full answer. One
still has either to prove that the answer has all the correct
discontinuities across all branch cuts, which was done in \Bernry,
or to prove that the basis of integrals is given entirely by
double boxes. Unfortunately, the basis of integrals is not known
at two loops.

For four gluons, even though the number of integration variables
is greater that the number of the delta-functions in a hepta-cut,
no integration has to be performed. We find that this is not the
case if the number of external gluons is greater than four. We
will see in the next section that already in the case of
five-gluon amplitude, the product of the corresponding tree-level
amplitudes does depend on the loop momenta and cannot be pulled
out of the integral.


\newsec{Octa-Cuts of Two-Loop Amplitudes}


In the introduction we distinguished between two different kinds
of double box scalar integrals. In the first class, the two boxes
share a propagator while in the second class they only share a
vertex, see fig. 1a and fig. 1b respectively. In this section, we
show how one can use octa-cuts to compute the coefficient of a
certain subset of the first class and the coefficients of all
integrals of the second class, which we called split double boxes.


\subsec{Double-Box Scalar Integrals}


Let us start with the double boxes that have seven propagators. We
will show that when at least one of the two boxes has two adjacent
three particle vertices then there is an extra propagator-like
singularity that can be cut. This produces one more delta-function
which together with the hepta-cut of the previous section
completely localizes the cut integral. Even though we concentrate
only on the planar configurations, exactly the same logic can be
applied for non-planar configurations as well. Consider an
arbitrary double-box configuration shown in fig. 5. The
corresponding hepta-cut integral is
\eqn\aone{ {\cal I} =\int {d^4p\over (2\pi)^4} {d^4q\over
(2\pi)^4} \delta(p^2) \delta ((p-k_1)^2) \delta ((p-k_1-k_2)^2)
\delta ((p+q+k_6)^2) \delta (q^2) \delta((q-k_5)^2) \delta
((q-k_4-k_5)^2).}
\ifig\figthree{An arbitrary double-box configuration.} {\epsfxsize=0.50\hsize\epsfbox{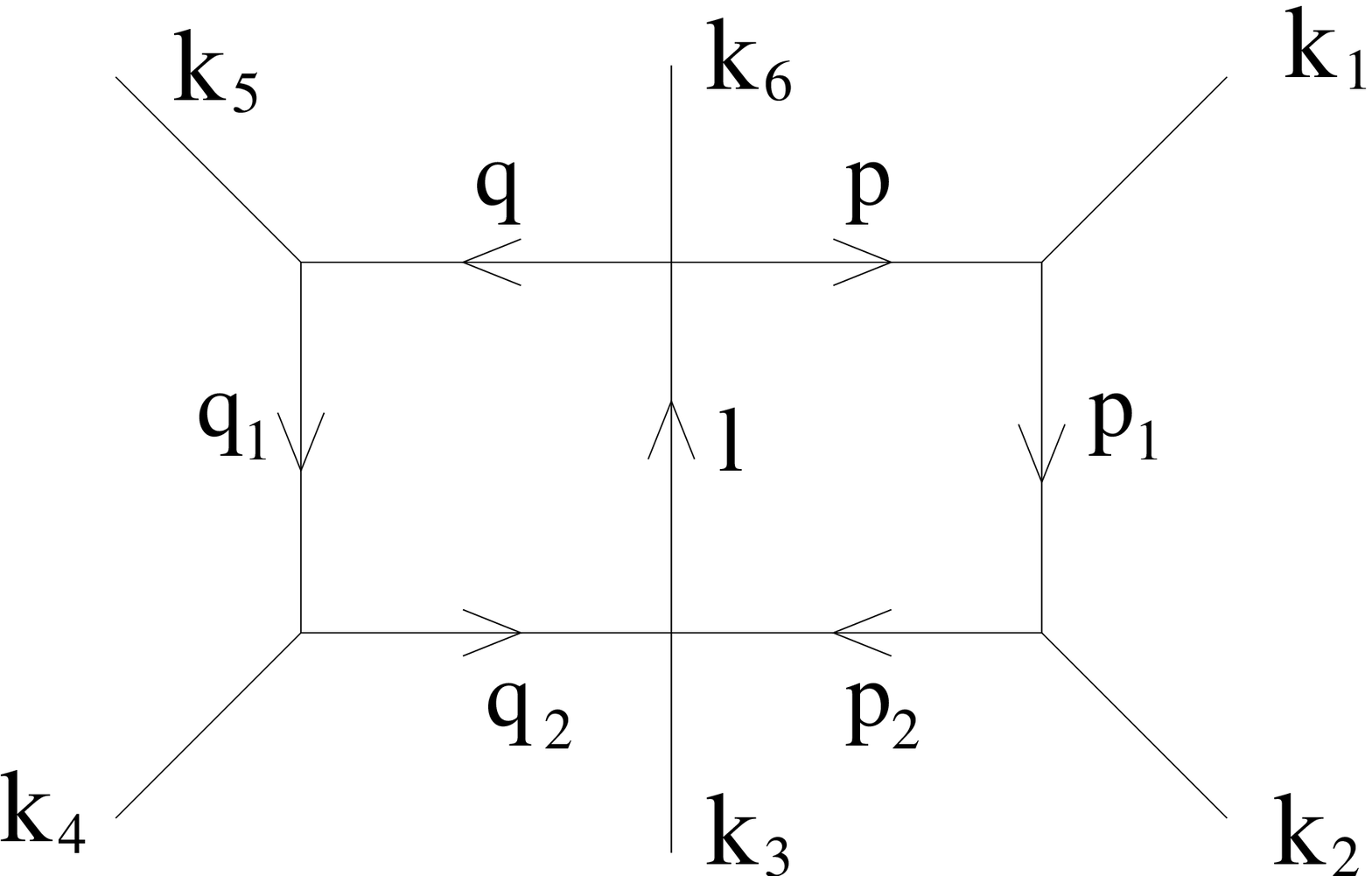}}
\noindent
Let us perform, for example, the $p$-integration. The integral
over $p$,
\eqn\atwo{
{\cal I}_p =\int d^4p \delta(p^2) \delta ((p-k_1)^2) \delta ((p-k_1-k_2)^2)
\delta ((p+q+k_6)^2),}
is localized and the answer is~\General
\eqn\athree{
{\cal I}_p={2 \over (k_1+k_2)^2 (k_1+k_6+q)^2\rho},}
where
\eqn\afour{\eqalign{
&\rho=\sqrt{1-2(\lambda_1+\lambda_2)+(\lambda_1-\lambda_2)^2}, \cr &
\lambda_1= {k_1^2 (k_3+k_4+k_5-q)^2 \over (k_1+k_2)^2 (k_1+k_6+q)^2}, \quad
\lambda_2= {k_2^2 (k_6+q)^2 \over (k_1+k_2)^2 (k_1+k_6+q)^2}.}}
The crucial observation is that when
\eqn\afive{\rho=1,}
${\cal I}$ acquires an extra propagator-type singularity, i.e.
$1/(k_1+k_6+q)^2$. We can formally cut the new propagator by
replacing it with a delta-function creating an eighth cut. In
other words, after performing the $p$-integration we end up with
following integral over $q$ (we omit the overall $q$-independent
factor)
\eqn\asix{
{\cal I}_q=\int d^4q \delta (q^2) \delta((q-k_4)^2) \delta ((q-k_3-k_4)^2)
{1 \over (k_1+k_6+q)^2}.}
This integral looks like a triple cut of the following effective box
\ifig\figfour{Effective box that arises after a quadruple cut is
used to localize the $p$ integral. The momentum flowing along the
uncut line is $q+k_1+k_6$.}
{\epsfxsize=0.50\hsize\epsfbox{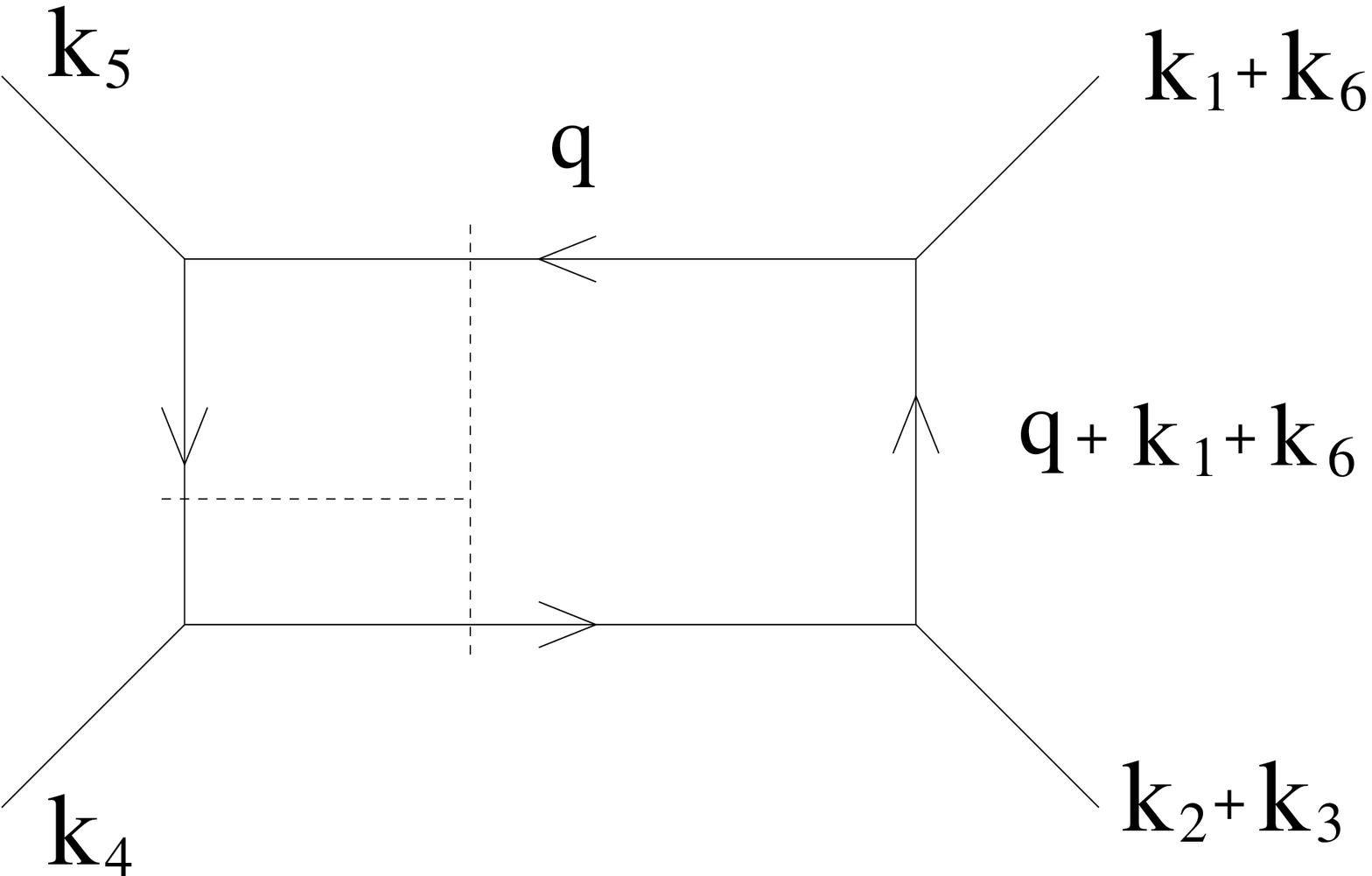}}
\noindent
Note that the momentum flowing along the uncut line is exactly
$q+k_1+k_6$. From this viewpoint it is natural to cut the
remaining propagator. Note that this procedure localizes the
$q$-integral. Then it is straightforward to write down the
coefficients of such double-box integrals. They are given by
\eqn\aseven{ c_{\alpha}= -{i \over |{\cal S}|}\sum_{h, J_1, J_2,
{\cal S}}
(n_{J_{1}}n_{J_{2}}A_{(1)}^{tree}A_{(2)}^{tree}A_{(3)}^{tree}A_{(4)}^{tree}A_{(5)}^{tree}A_{(6)}^{tree})_h,}
where the sum over $h$ is the sum over all helicity
configurations, the sums over $J_1$ and $J_2$ are the sums over
all particles that can propagate in both loops, ${\cal S}$ is the
set of all solutions for the internal lines of the following
system of equations
\eqn\aeight{\eqalign{ & p^2 =0, \quad (p-k_1)^2=0, \quad
(p-k_1-k_2)^2=0, \quad (p+q+k_6)^2=0, \cr & q^2=0, \quad (q-k_5)^2
=0, \quad (q-k_4-k_5)^2=0, \quad (k_1+k_6+q)^2=0,}}
and $|{\cal S}|$ is the number of solutions. This expression is
analogous to the formula for one-loop coefficients of box
integrals~\General. It is important to remember that this
discussion is valid if
\eqn\anine{
\rho=1, \quad \lambda_1=\lambda_2=0.}
Otherwise, the singularity $1/(k_1+k_6+q)^2$ will be replaced by a
more complicated one which is not propagator-like, as it can
easily be seen from eq.~\afour. The conditions given in~\anine\
are satisfied if a given box has two adjacent three-particle
vertices. It easy to check that this is always the case if the
number of gluons is less than seven. This means that every
double-box coefficient of any gluon amplitude with less than seven
external lines is given by eq.~\aseven. The first double-box
configuration where eq.~\anine\ is not satisfied appears when the
number of external gluons is seven and is shown in fig. 7.
\ifig\figfive{The simplest double-box configuration for which the
conditions in~\anine\ are not satisfied.}
{\epsfxsize=0.50\hsize\epsfbox{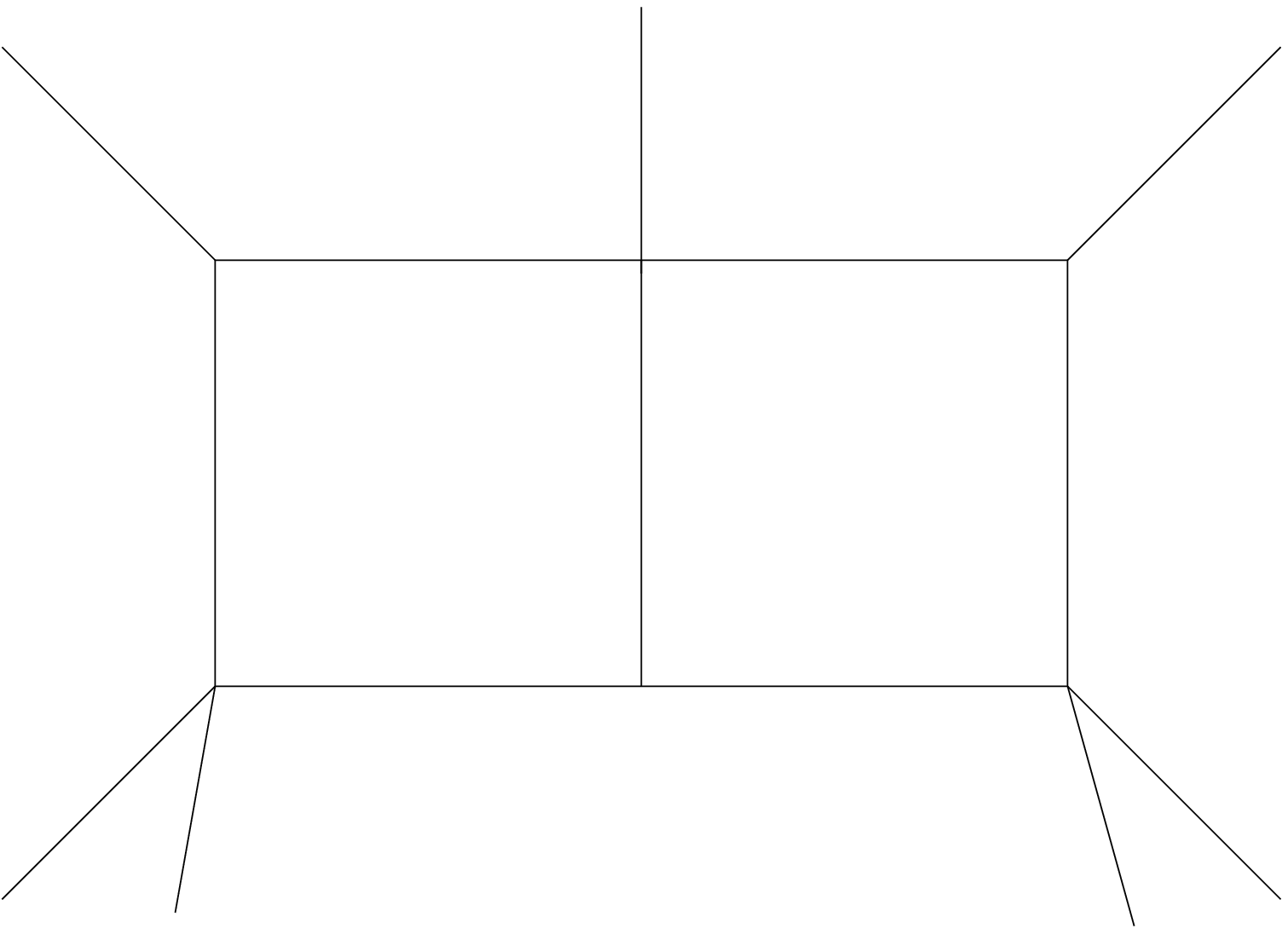}}
\noindent
However, even if the number of external gluons is greater than
six, there are double-box configurations for which eq.~\anine\ is
satisfied. In such cases the eighth cut still exists and
eq.~\aseven\ is still correct.

In fact, eq.~\aseven\ requires some additional explanations. Note
that existence of the effective box in fig. 6 implies that either
the momentum $l$ or the momentum $p_1$ in fig. 5 vanishes. In
Minkowski space, this would mean that some tree level amplitudes
in eq.~\aseven\ vanish. Moreover, in Minkowski space, the system
of equations~\aeight\ does not have solutions, which means that we
cannot see the singularities under consideration. Therefore, it is
not surprising that eq.~\aseven, at least naively, is meaningless
in Minkowski space. In order to see the new kind of singularities,
we have to analytically continue all momenta to signature
$(- - + +)$. But in signature $(- - + +)$, the statement that a
tree amplitude vanishes when one of the incoming or outgoing
momentum vanishes is not correct. Each tree amplitude is
constructed by using spinors. When one of the incoming or outgoing
$(- - + +)$ momenta vanishes, it is impossible to determine its
spinors components even up to rescaling. This leaves the amplitude
undetermined. For example, assume that we have a helicity
configuration containing a three-gluon amplitude $A(p^-, p_1^-,
k_1^+)$. It is given by
\eqn\insertone{
A(p^-, p_1^-, k_1^+)= {\vev{p_1~p}^3 \over \vev{p~k_1} \vev{k_1~p_1}}.}
If $p_1$ vanishes, the spinor $\lambda_{p_1}$ cannot be uniquely
determined. In fact, $\lambda_{p_1}$ is not uniquely defined even
for non-zero $p_1$ as it is defined up to rescaling. However, when
$p_1=0$ the freedom in not being able to determine $\lambda_{p_1}$
becomes much larger. One can always say that $p_1=0$ implies that
$\lt_{p_1}=0$ and $\lambda_{p_1}$ is arbitrary. This means that
$A(p^-, p_1^-, k_1^+)$ becomes arbitrary. Therefore, the numerator
in eq.~\aseven\ is a discontinuous function of momenta and we have
to give a prescription on how to define it as $l$ or $p_1$ goes to
zero. The natural way to define it is as follows. Consider first the
loop with momentum $p$. Let $A_{(1)}^{tree}, A_{(2)}^{tree},
A_{(3)}^{tree}$ and $A_{(4)}^{tree}$ be the four tree amplitudes
which depend on $p$. Assuming that they are all non-zero, we can
solve the first four $p$-dependent equations in~\aeight\ to
determine $p$ as a function of the external momenta and $q$ and then
evaluate the product
$A_{(1)}^{tree}A_{(2)}^{tree}A_{(3)}^{tree}A_{(4)}^{tree}$ on these
solutions. We claim that this product can be simplified in such a
way that it is a well-defined function when the constraint
$(k_1+k_6+q)^2=0$ is imposed. Below, we will present a few examples
that show that this is indeed the case. Having found the product
$A_{(1)}^{tree}A_{(2)}^{tree}A_{(3)}^{tree}A_{(4)}^{tree}$ as a
function of the external momenta and $q$, we then multiply it by the
remaining two tree amplitudes $A_{(5)}^{tree}$ and $A_{(6)}^{tree}$
and evaluate the product on the solution of the remaining four
equations in~\aeight. We propose this as a method for calculating
double-box coefficients provided conditions~\anine\ are fulfilled.

\bigskip

\noindent{\it A Subtlety}

\bigskip

There is one important subtlety we have to discuss before
presenting examples. Consider a helicity configuration with two
adjacent three-particle vertices, one of which depends only on
internal momenta and the other one depends on both internal and
external momenta, with both vertices having the same helicity
configuration. For example, consider the configuration in fig. 8.
\ifig\figsix{This helicity configuration is non-zero only if $\l_q
\sim \l_1$.} {\epsfxsize=0.50\hsize\epsfbox{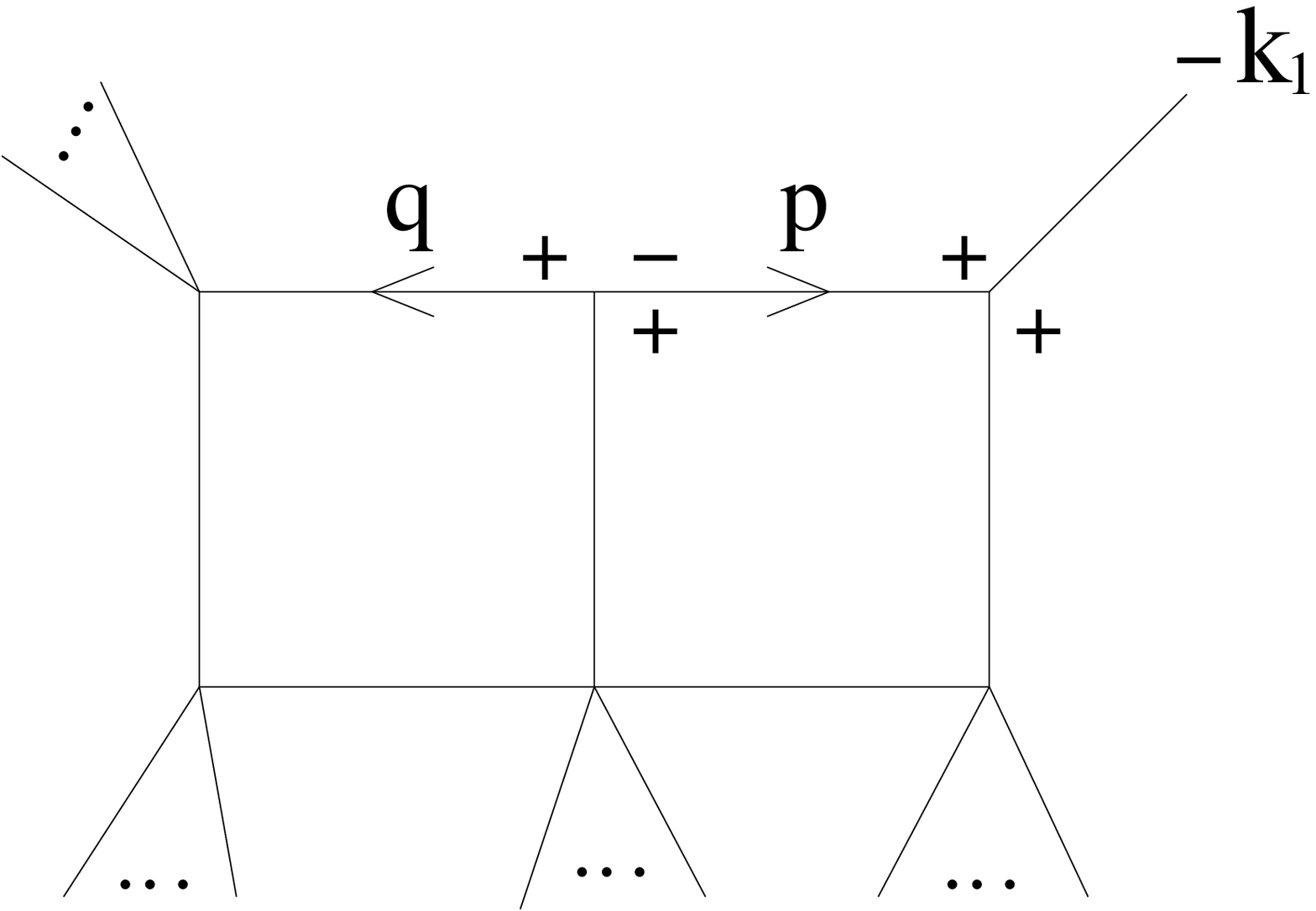}}
\noindent
This configuration is non-zero only if $\l_q \sim \l_1$.
Therefore, the integral over $p$
\eqn\atena{ \int d^4p \delta(p^2) \delta ((p-k_1)^2) \delta
((p-k_1-k_2)^2) \delta ((p+q)^2)
A_{(1)}^{tree}A_{(2)}^{tree}A_{(3)}^{tree}A_{(4)}^{tree}}
must be proportional to $\delta((k_1+q)^2)$. In other words, the
integral lacks the extra propagator-like singularity and therefore
it does not contribute to the octa-cut.

\subsec{Split Double-Box Scalar Integrals}

When the number of gluons is greater than five, a new kind of
double box integrals can appear. These were introduced in the
introduction in fig. 1b. For the reader's convenience we depict
them again in fig. 9. This double box scalar integrals are such
that the two boxes only share a vertex and not a propagator. This
is why we will call them split double boxes.
\ifig\figsixteen{Generic split double box configuration.}
{\epsfxsize=0.40\hsize\epsfbox{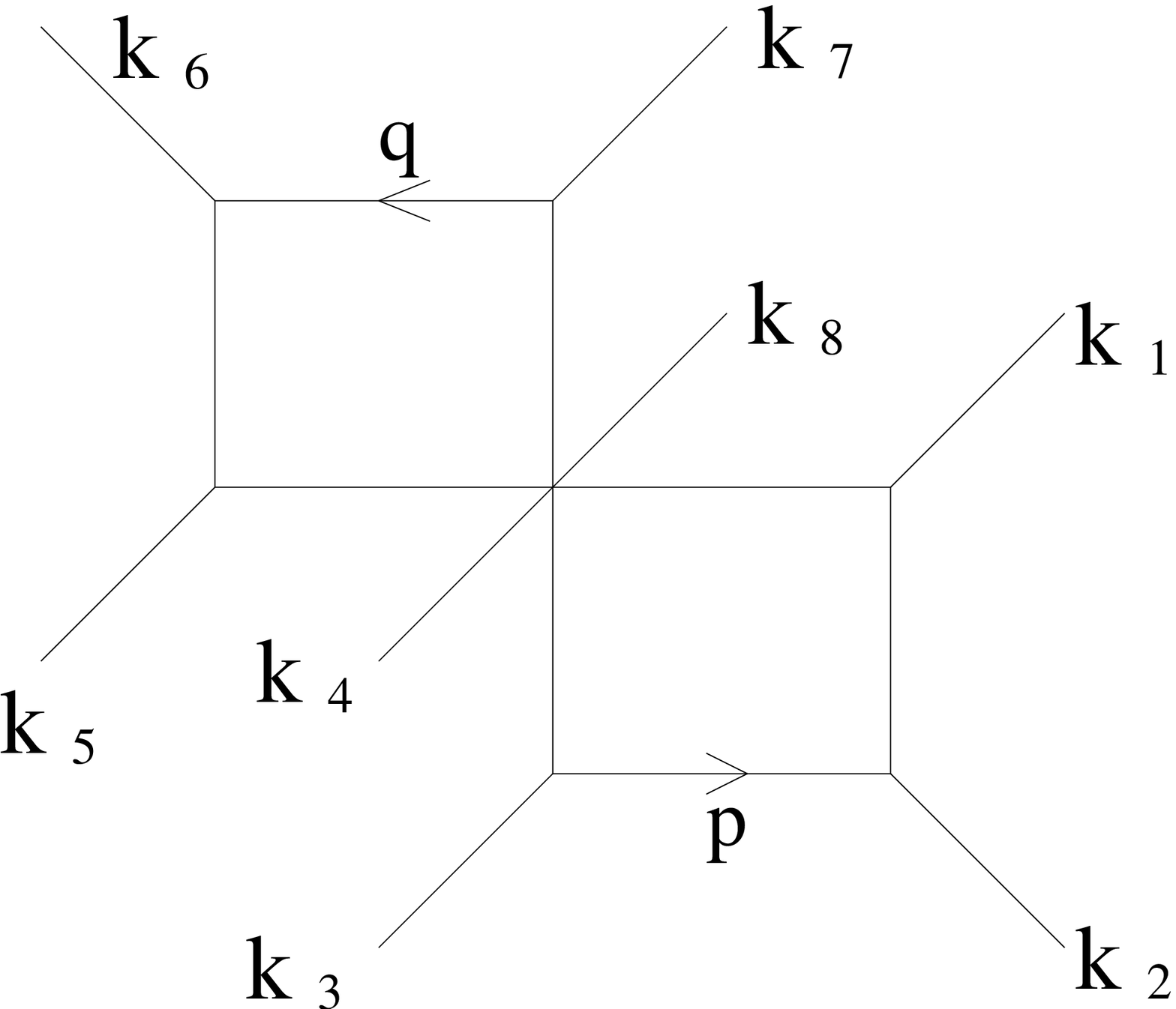}}
\noindent The coefficients of the split double boxes are easy to
compute. Since they have eight propagators and the two loop
integrations are completely independent, it is straightforward to
consider two quadruple cuts or equivalently an honest octa-cut.
This produces eight delta-functions that localize both loop
integrals. Let us see this in more detail. The quadruple cut in
the $q$-loop fixes $q$ to be a solution to the following
equations,
\eqn\newqq{q^2=0, ~ (q-k_6)^2=0,~ (q-k_5-k_6)^2=0,~ (q+k_7)^2=0,}
while a quadruple cut in the $p$-loop fixes $p$ to be a solution
of
\eqn\newpp{p^2=0, ~ (p-k_2)^2=0, ~ (p-k_1-k_2)^2=0, ~ (p+k_3)^2 =
0. }
Each set of equations has two solutions. The coefficient of a
split double-box scalar integral is then given by
\eqn\splitco{ c = {1 \over 4}\sum_{h,J_1,J_2,{\cal
S}}(n_{J_1}n_{J_2}
A_{(1)}^{tree}A_{(2)}^{tree}A_{(3)}^{tree}A_{(4)}^{tree}A_{(5)}^{tree}A_{(6)}^{tree}A_{(7)}^{tree})_h
}
where we have used that the number of solutions is $4$.


\newsec{Examples}


In this section we consider several examples of coefficients
calculated by using octa-cuts. All of them are coefficients of
scalar double boxes with seven propagators. We consider four-,
five- and six-gluon amplitudes. For six-gluons we study a non-MHV
amplitude with adjacent negative helicity gluons.


\subsec{Four-Gluon Amplitude $A^{2-loop}(-, -, +, +)$ Revisited}


As a first example, let us reconsider the octa-cut of the
four-gluon amplitude from section 3 in the $s$-channel. The octa
cut in the $t$-channel is analogous. The additional propagator
that we cut is ${1\over (q+K_1)^2}$. See fig. 3 for notation.
Taking into account the subtlety in the previous section, there
are only four helicity configurations that contribute. All of them
give the same answer $-A_{4}^{tree}s^2 t$. On the other hand, the
number of solutions to eqs.~\aeight\ can be shown to be four. This
gives $c_s=-A_{4}^{tree}s^2 t$ as in eq.~\enine. Note that the
product of the four tree level amplitudes depending on the
internal momentum $p$ is given by $\vev{1~2}^2 [q~q_2]^2$ (see
eq.~\etwelve). This expression does not have any ambiguity in the
presence of the eighth delta-function $\delta ((q+K_1)^2)$.


\subsec{Five-Gluon Amplitude $A^{2-loop}(-, -, +, +, +)$}


As a next example, let us calculate the coefficient of the following
five-gluon double-box configuration.
\ifig\figseven{A double box integral of the five-gluon amplitude
$A(1^-,2^-,3^+,4^+,5^+)$.}
{\epsfxsize=0.50\hsize\epsfbox{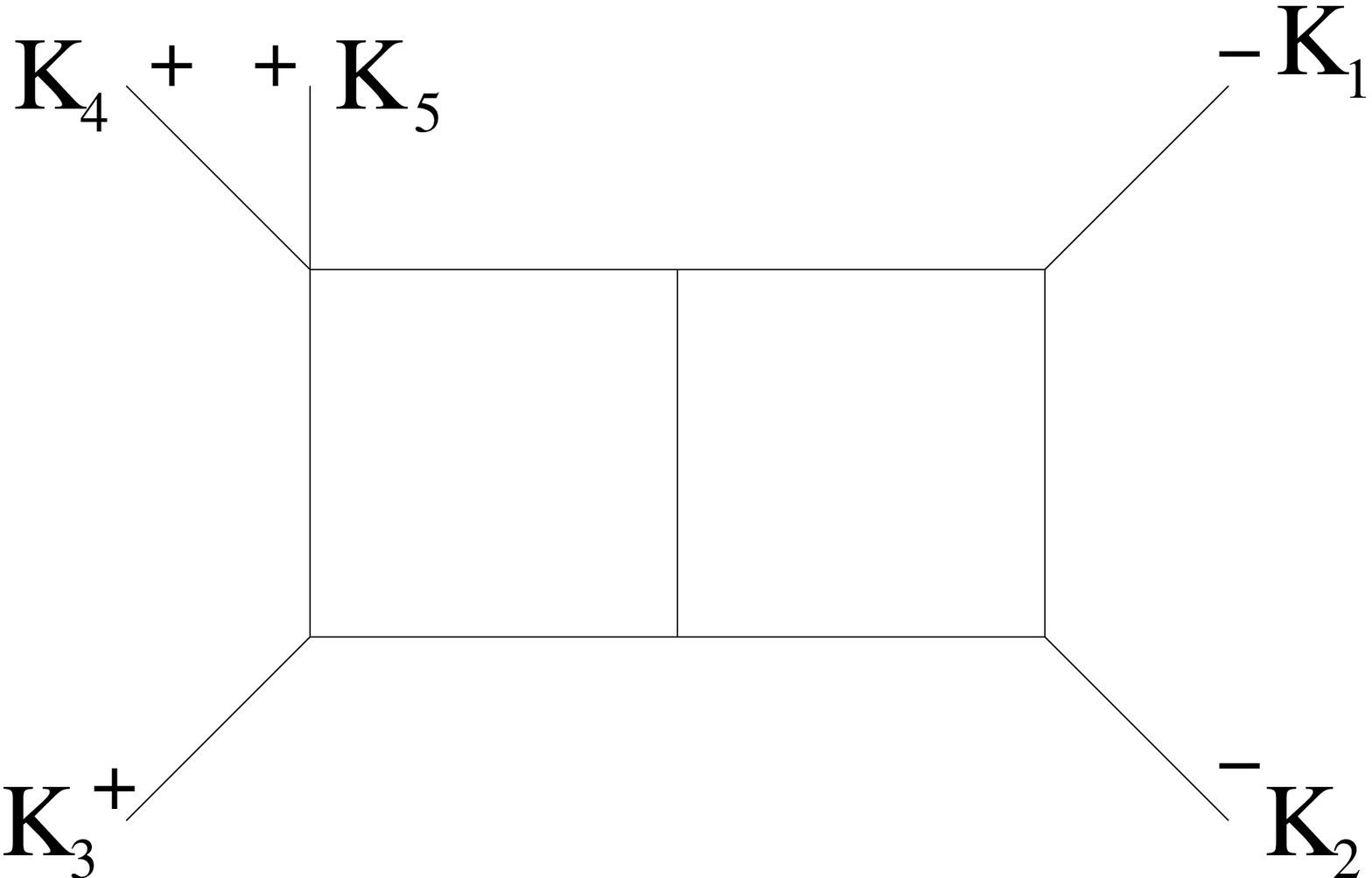}}
\noindent
In this case, there are two helicity configurations to consider.
Both of them can be shown to give the same contribution. We will consider
the helicity configuration shown in fig. 11.
\ifig\figeight{One of the two possible helicity configurations
contributing to the coefficient of the integral of fig. 10.}
{\epsfxsize=0.50\hsize\epsfbox{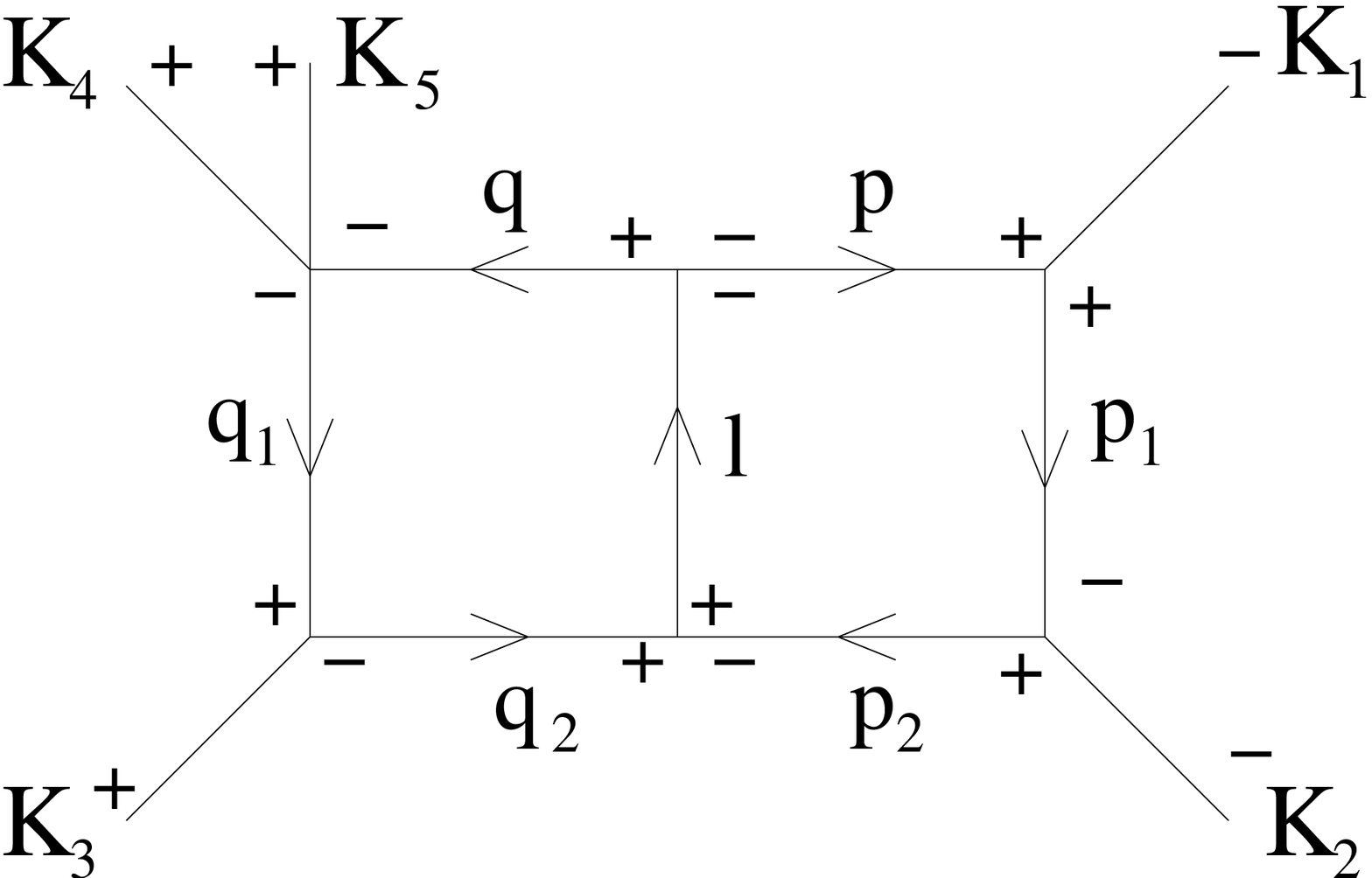}}
\noindent Note that only gluons can propagate in both loops. The
product of the six tree level amplitudes is as follows
\eqn\aten{
\vev{1~2}^2[q~q_2]^2
{[3~q_1]^3\over [q_1~q_2][q_2~3]}
{\vev{q~q_1}^3 \over \vev{q_1~4}\vev{4~5}\vev{5~q}},}
where the first two factors come from the four vertices on the
right. The computation leading to the first two factors was done
in the previous section in eq.~\etwelve. By using momentum
conservation along the lines of eq.~\eeleven, eq.~\aten\ can be
reduced to
\eqn\aeleven{
{\vev{1~2}^2 s^2 \over \vev{3~4}\vev{4~5}}
\left([5~3]+[4~3] {\vev{q~4} \over \vev{q~5}}\right),}
Now we impose the delta-function $\delta((K_1+q)^2)$.
It has two possible solutions, $\l_q \sim \l_1$ or $\lt_q \sim \lt_1$.
It is not hard to show that if we choose $\lt_q \sim \lt_1$ then the expression in eq.~\aeleven\ vanishes.
Therefore, the only relevant solution is $\l_q \sim \l_1$.
Then the octa-cut becomes
\eqn\atwelve{
{\vev{1~2}^2 s^2 \over \vev{3~4}\vev{4~5}}
\left([5~3]+[4~3] {\vev{1~4} \over \vev{1~5}}\right)\int d\mu.}
Taking into account that the system of equations~\aeight\ has four solutions, we
find that the corresponding coefficient is
\eqn\athirteen{ c^{(5)}_1= -{2i \over 4}{\vev{1~2}^2 s^2 \over
\vev{3~4}\vev{4~5}} \left([5~3]+[4~3] {\vev{1~4} \over
\vev{1~5}}\right)=-{1 \over 2}A_5^{tree}s^2t,}
where $s=(K_1+K_2)^2$ and $t=(K_2+K_3)^2$ and $A_5^{tree}$ is the
tree-level five-gluon amplitude
\eqn\afourteen{
A_5^{tree}=i
{\vev{1~2}^3 \over \vev{2~3}\vev{3~4}\vev{4~5}\vev{5~1}}.}

Let us consider one more example. Let us calculate
the coefficient of the
five-gluon double-box configuration shown in fig. 12.
\ifig\fignine{Second example of a double-box configuration in the
five-gluon amplitude $A(1^-,2^-,3^+,4^+,5^+)$.}
{\epsfxsize=0.50\hsize\epsfbox{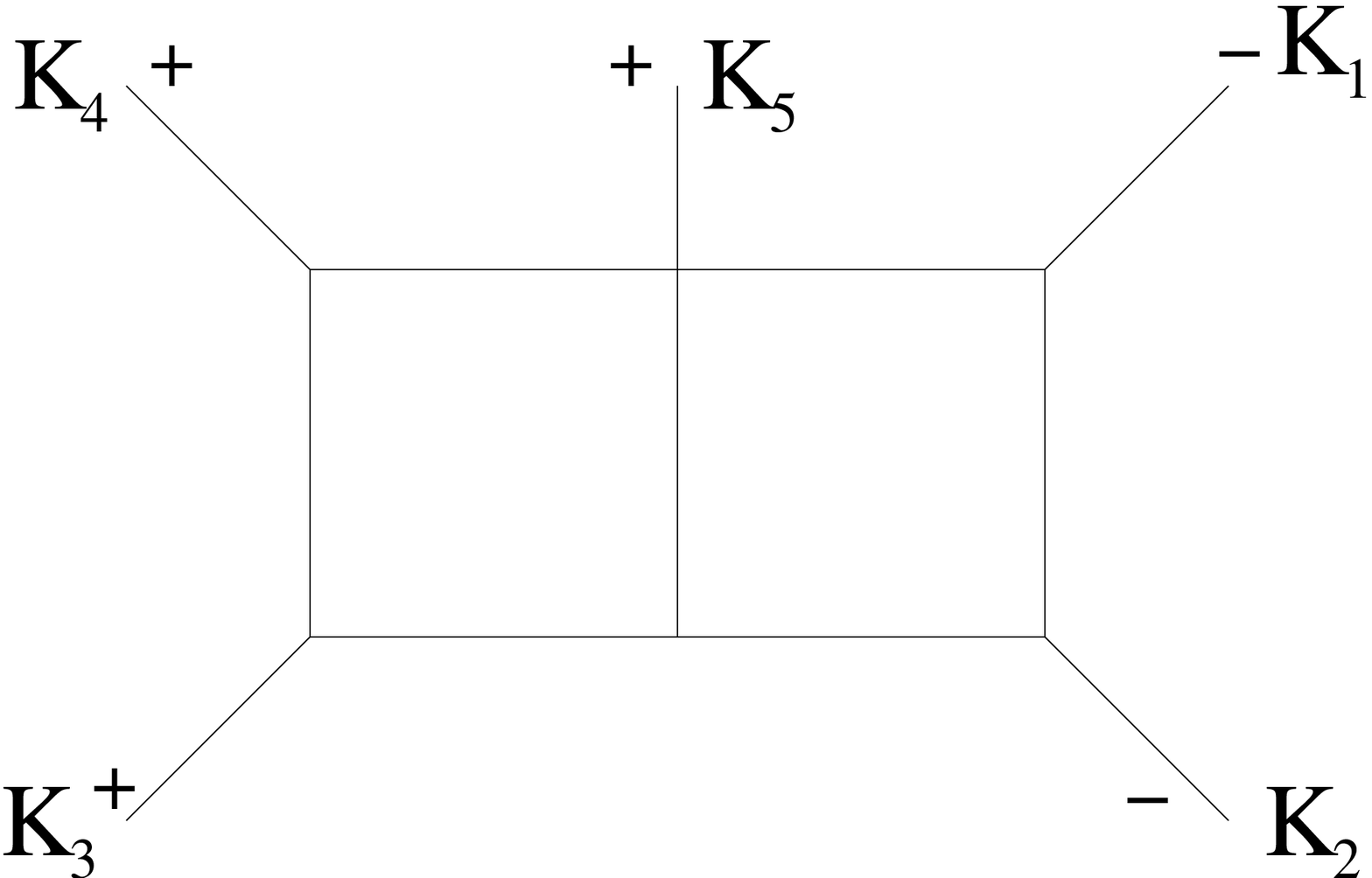}}
\noindent
There are two helicity configurations to consider. However, one of
them gives the zero answer. The non-zero contribution comes from the
helicity configuration shown in fig. 13.
\ifig\figten{The only non-vanishing helicity configuration
contributing to the coefficient of the double box integral of fig.
12.} {\epsfxsize=0.50\hsize\epsfbox{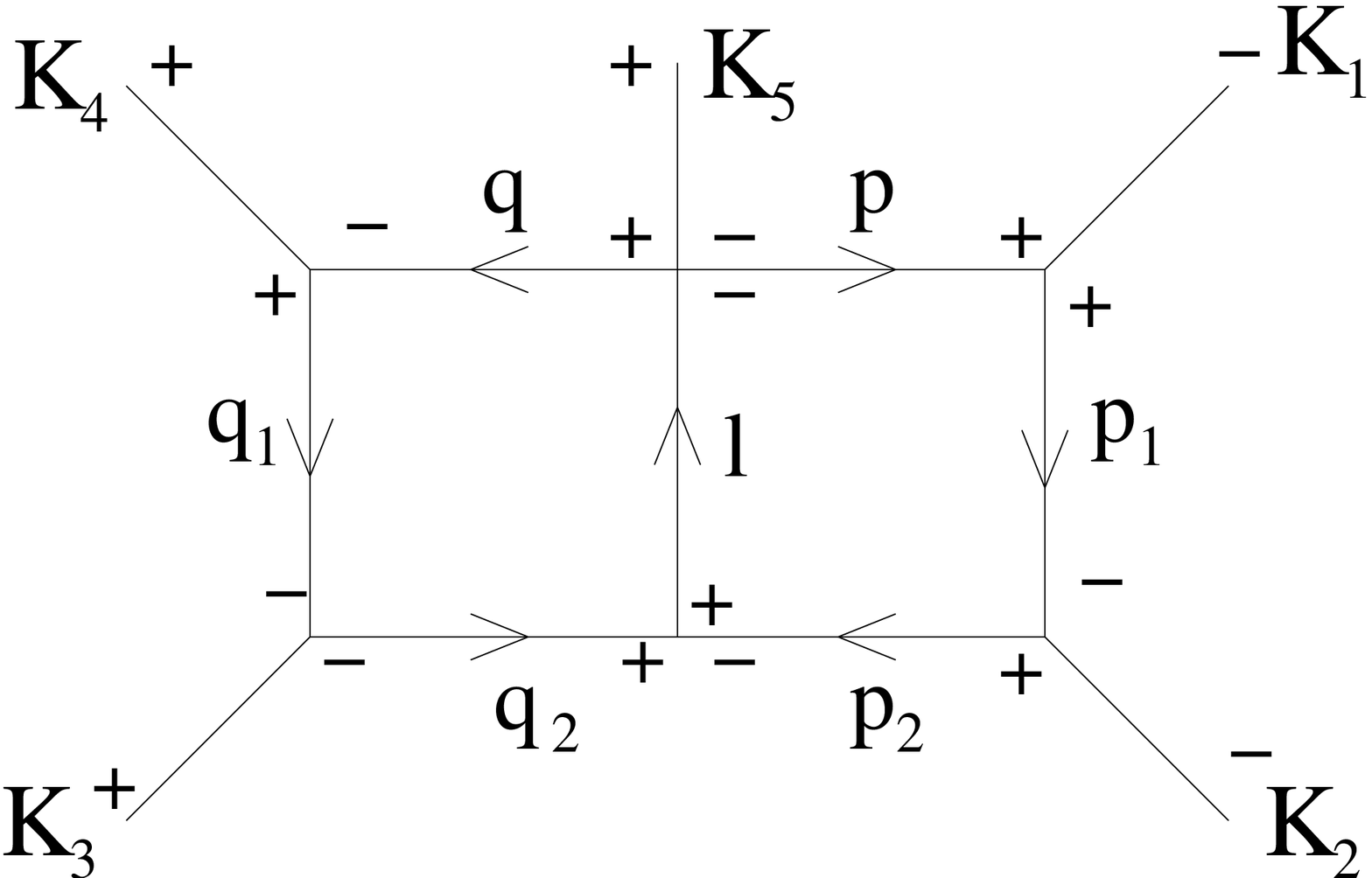}}
\noindent
Note that only gluons can propagate in both loops. The number of
solutions to eqs.~\aeight\ can be shown to be two. Then the
corresponding coefficient is given by
\eqn\afifteen{\eqalign{ c^{(5)}_2= & -{i \over 2} {\vev{q_1~q_2}^3
\over \vev{q_2~3} \vev{3~q_1}} {[q_1~4]^3 \over [4~q][q~q_1]}
{\vev{p~l}^3 \over \vev{l~q} \vev{q~5}\vev{5~p}} {[q_2~l]^3 \over
[l~p_2][p_2~q_2]}\cr & {\vev{p_1~2}^3 \over \vev{2~p_2}
\vev{p_2~p_1}} {[p_1~p]^3 \over [p~1][2~p_1]}.}}
Using momentum conservation and the fact that $\lambda_q \sim
\lambda_4$, eq~\afifteen\ can be simplified to give
\eqn\atwenty{
c_2^{(5)}=-{1 \over 2} A_5^{tree} stu,}
where $u=(K_3+K_4)^2$. All other double-box coefficients of the
five-gluon amplitude can be found by analogous calculations.


\subsec{Six-Gluon Amplitude $A^{2-loop}(-, -, -, +, +, +)$}


As our next example, let us calculate the coefficient of the
six-gluon next-to-MHV double-box configuration shown in fig. 14.
The additional singularity that we cut is again ${1\over
(q+K_1)^2}$. There are two helicity configurations to consider,
both giving the same answer. In both of them only gluons can
propagate in both loops. Let us describe the calculation of the
one depicted in fig. 15.
\ifig\figeleven{A double-box integral of the six-gluon non-MHV
amplitude $A(1^-,2^-,3^-,4^+,5^+,6^+)$.}
{\epsfxsize=0.50\hsize\epsfbox{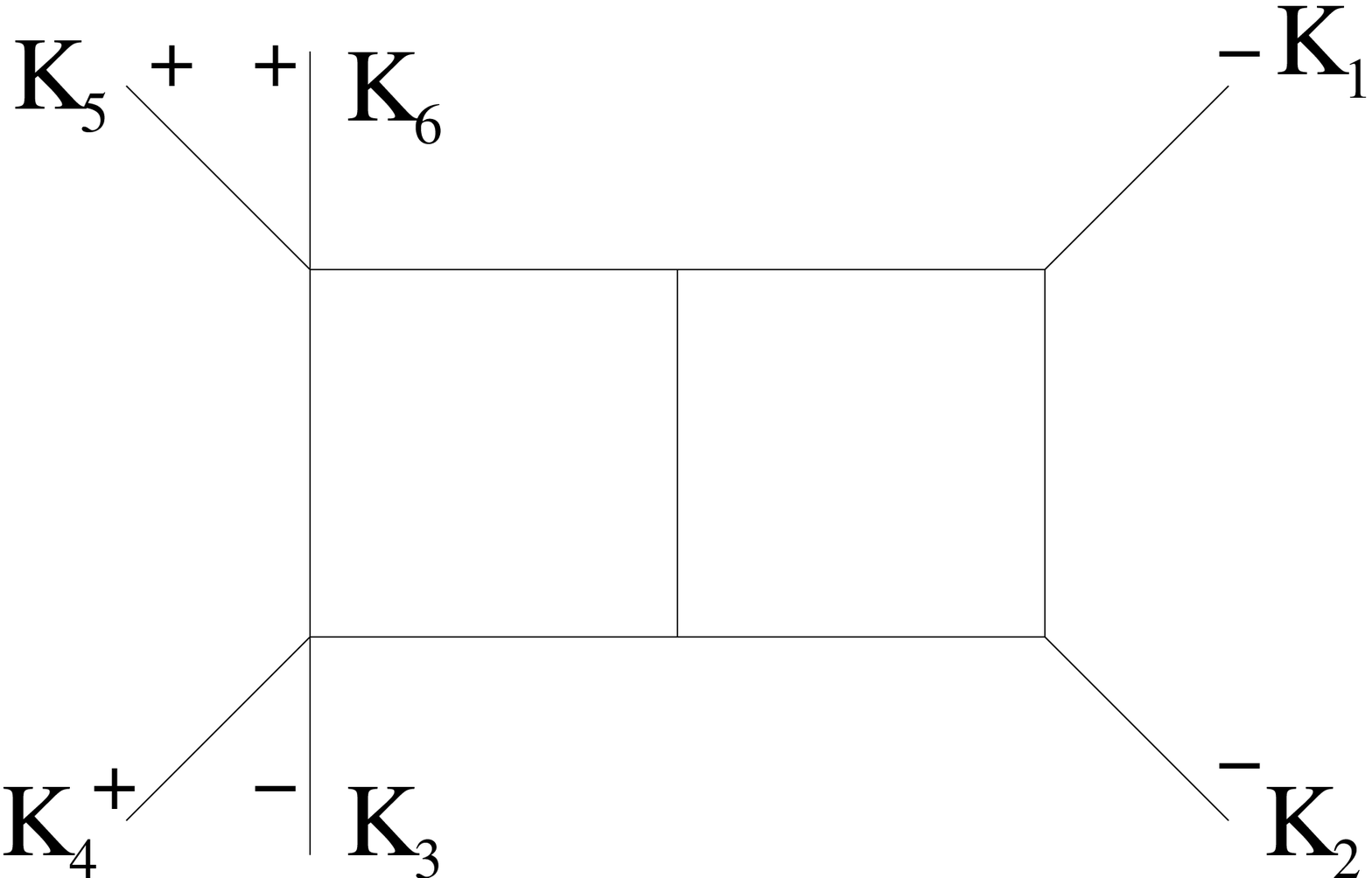}}
\ifig\figtwelve{One of the two helicity configurations
contributing to the coefficient of the scalar double box of
fig.14.} {\epsfxsize=0.50\hsize\epsfbox{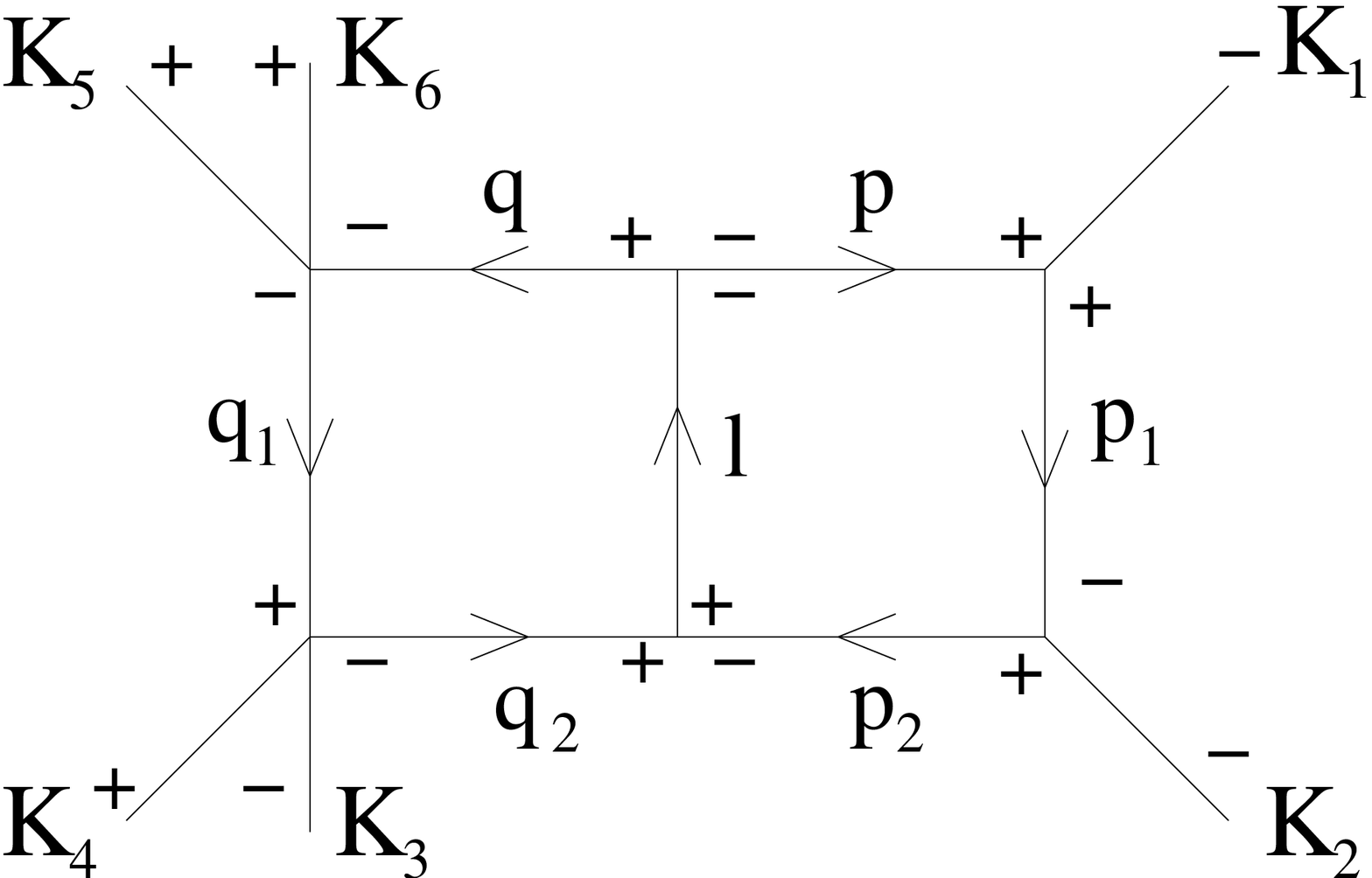}}
\noindent
The product of the four
$p$-dependent tree amplitudes gives (see eq.~\etwelve)
\eqn\atwentyone{
\vev{1~2}^2 [q~q_2]^2.}
Then the numerator of eq.~\aseven\ becomes
\eqn\atwentytwo{\eqalign{ & - 2 i
(A_{(1)}^{tree}A_{(2)}^{tree}A_{(3)}^{tree}A_{(4)}^{tree}A_{(5)}^{tree}A_{(6)}^{tree})=\cr
& \vev{1~2}^2 [q~q_2]^2 {\vev{q_2~3}^3 \over
\vev{3~4}\vev{4~q_1}\vev{q_1~q_2}} {[5~6]^3 \over
[6~q][q~q_1][q_1~5]},}}
where the factor of two comes from two helicity configurations.
Using momentum conservation similar to eq.~\etwelve, eq.~\atwentytwo\
can be simplified as follows
\eqn\atwentythree{
i{\vev{1~2}^2[5~6]^3 \gb{3|5+6|q}^2 \vev{q_2~3} \over \vev{3~4}[6~q]
\gb{4|5+6|q}\gb{q_2|3+4|5}}.}
Now we impose the last condition $(q+K_1)^2=0$. This equation has
two solutions. We can either have $\lt_q \sim \lt_1$ or $\l_q \sim
\l_1$. Both solutions give non-zero contributions. The first
solution yields
\eqn\atwentyfour{ i{\vev{1~2}^2[5~6]^3 \gb{3|5+6|1}^2 \vev{2~3}
\over \vev{3~4}[6~1] \gb{4|5+6|1}\gb{2|3+4|5}}}
while the second solution yields
\eqn\atwentyfive{
i{\vev{1~2}^2[5~6]^3\vev{5~6}[4~2]^2 \gb{3|(1+2)\cdot(5+6)\cdot(3+4)|2}
\over [2~3]\gb{5|6+1|2}[5|(3+4)\cdot(1+2)\cdot(5+6)\cdot(3+4)|2]}.}
Taking into account that the system~\aeight\ in this case has four solutions,
the double-box coefficient becomes
\eqn\atwentysix{\eqalign{
c_1^{(6)}= & {i \over 2} \left( {\vev{1~2}^2[5~6]^3 \gb{3|5+6|1}^2 \vev{2~3} \over \vev{3~4}[6~1]
\gb{4|5+6|1}\gb{2|3+4|5}} + \right. \cr & \left.
{\vev{1~2}^2[5~6]^3\vev{5~6}[4~2]^2 \gb{3|(1+2)\cdot(5+6)\cdot(3+4)|2}
\over [2~3]\gb{5|6+1|2}[5|(3+4)\cdot(1+2)\cdot(5+6)\cdot(3+4)|2]} \right).}}

Let us consider one more example. Let us calculate the coefficient
of the six-gluon double-box configuration shown in fig. 16.
\ifig\figthirteen{Second example of a six-gluon double-box
integral of the six-gluon non-MHV amplitude
$A(1^-,2^-,3^-,4^+,5^+,6^+)$.}
{\epsfxsize=0.50\hsize\epsfbox{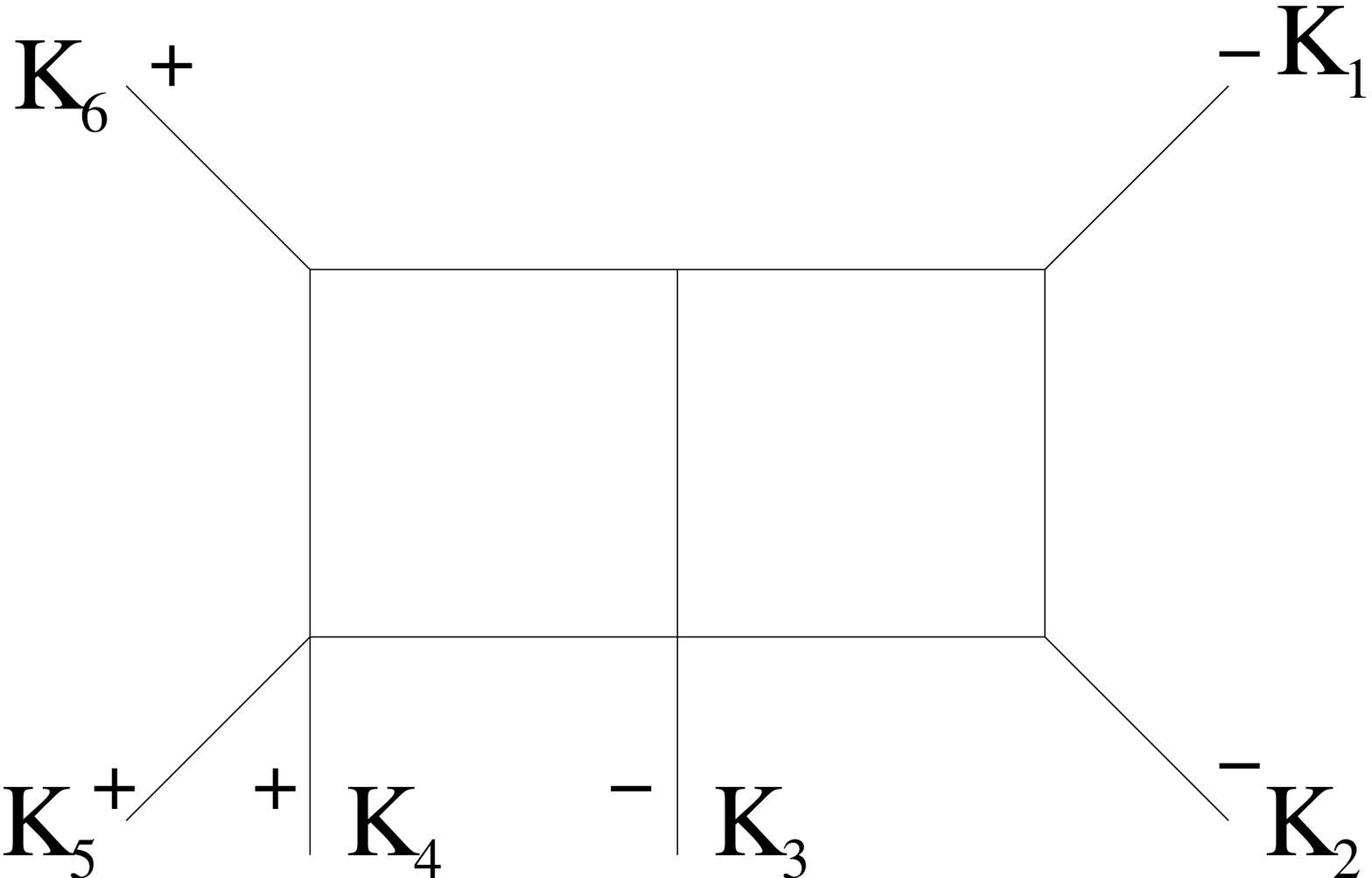}}
%
%
\ifig\figfourteen{The only non-vanishing helicity configuration
contributing to the coefficient of the scalar double-box integral
of fig.16.} {\epsfxsize=0.50\hsize\epsfbox{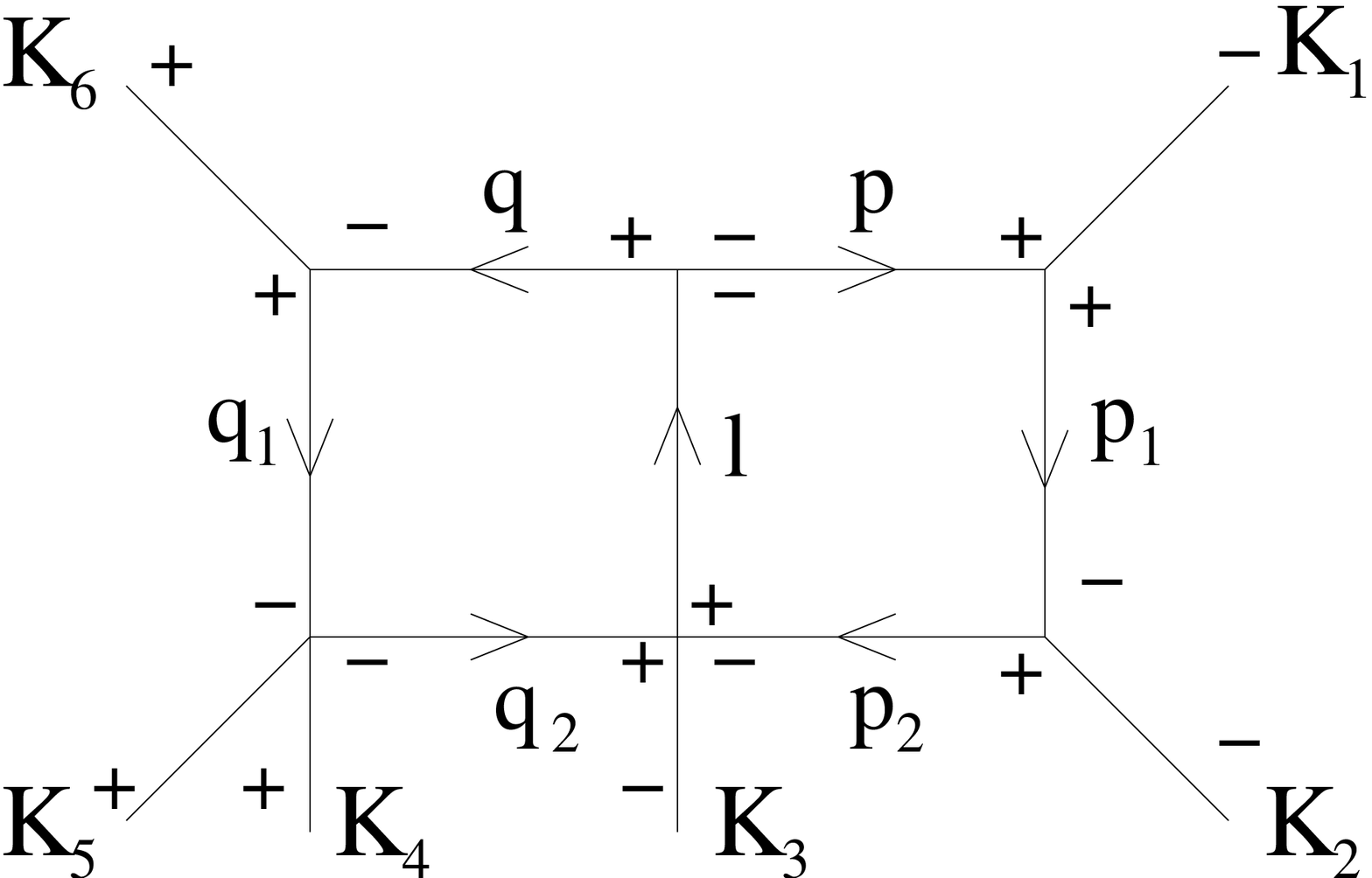}}
\noindent
In this case, there is only
one helicity configuration contributing to the octa-cut. It is
shown in fig. 17.
Only gluons can propagate in both loops. The system of
equations~\aeight\ has two solutions. Therefore, the corresponding
double-box coefficient is given by
\eqn\atwentyseven{\eqalign{ c^{(6)}_2= & -{i \over 2} {[p_1~p]^3
\over [p_1~1][1~p_1]} {\vev{p_1~2}^3 \over \vev{2~p_2}
\vev{p_2~p_1}} {\vev{p~l}^3 \over \vev{l~q}\vev{q~p}} {[q_2~l]^3
\over [l~p_2][p_2~3][3~q_2]} \times \cr & {[q_1~6]^3 \over
[6~q][q~q_1]}{\vev{q_1~q_2}^3 \over \vev{q_2~4} \vev{4~5}
\vev{5~q_1}}.}}
By using the first seven equations in~\aeight, we can simplify~\atwentyseven\
as follows
\eqn\atwentyeight{
c^{(6)}_2=-{i \over 2} {u^3 s \gb{1|q|6} \over
[1~2][2~3]\vev{4~5}\vev{5~6} \gb{4|q-K_{4}^{[3]}|3}},}
where
\eqn\atwentynine{
u=(K_4+K_5+K_6)^2, \quad s=(K_1+K_2)^2.}
Now we consider the last equation $(q+K_1)^2=0$. From fig. 17, it follows
that $\l_q$ has to be proportional to $\l_6$. Therefore, $\lt_q$ has to be
proportional to $\lt_1$. Using momentum conservation, one can find that
\eqn\athirty{
q=-{u \over \gb{6|4+5|1}} \l_6 \lt_1.}
Substituting this into eq.~\atwentynine,
we obtain
\eqn\athirtyone{
c^{(6)}_2 =-{i \over 2}
{u^4 st \over [1~2][2~3]\vev{4~5}\vev{5~6} \gb{4|5+6|1} \gb{6|4+5|3}},}
where
\eqn\athirtytwo{
t=(K_1+K_6)^2.}
All other double-box coefficients can be computed by similar calculations.

The calculation of the coefficient $c^{(6)}_2$
can be generalized for the configuration considered in fig. 18.
\ifig\figfifteen{An infinite family of $n$-gluon double box scalar
integrals.} {\epsfxsize=0.50\hsize\epsfbox{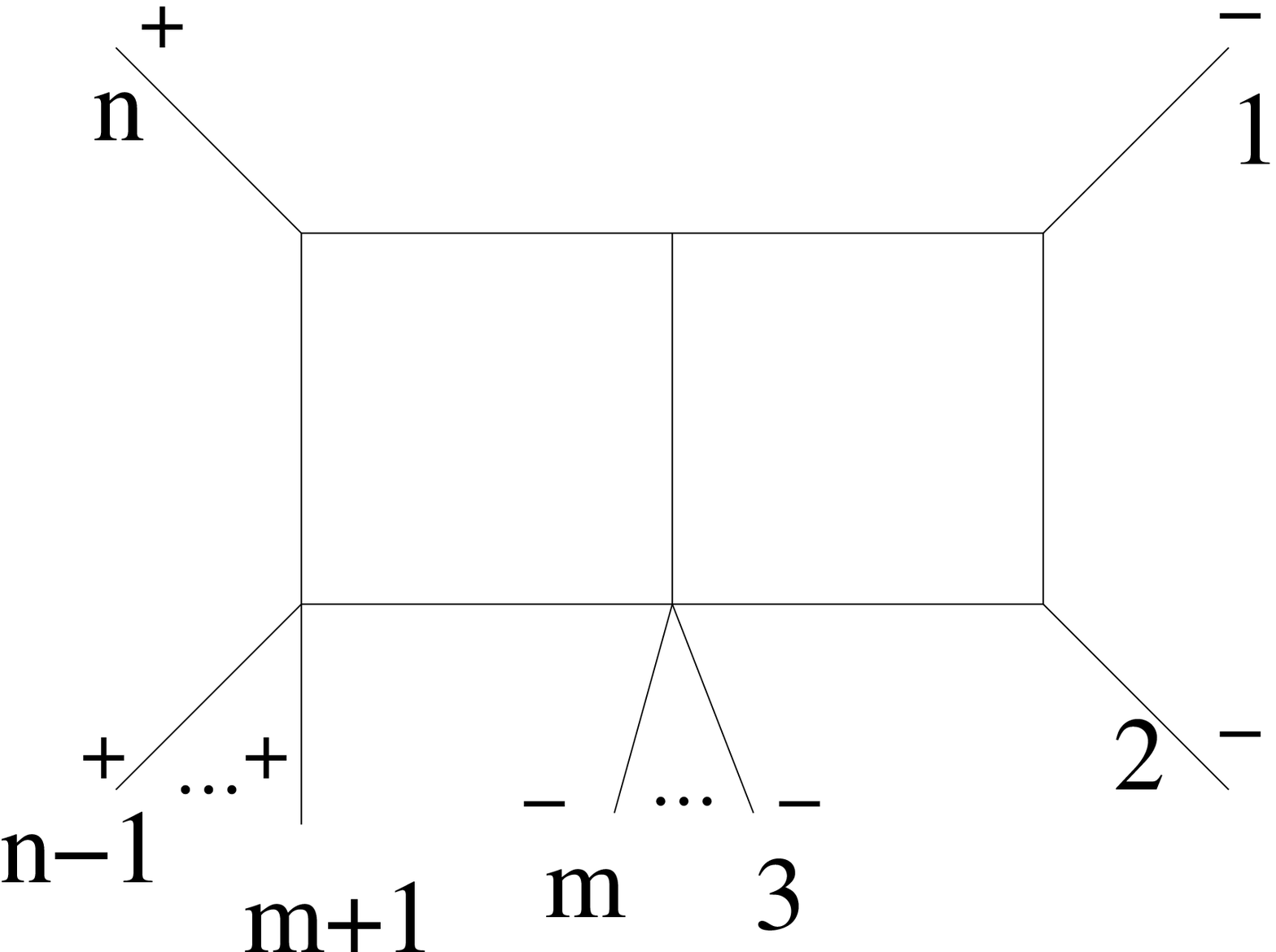}}
\noindent There is only one helicity configuration that
contributes to the octa-cut. A similar calculation gives
\eqn\athirtythree{\eqalign{
c^{(n)} =&-{i \over 2}
{u^4 st \over [1~2] [2~3] \dots [m-1~m] \vev{m+1~m+2}\dots \vev{n-1~n}} \times \cr &
{1 \over \gb{m+1|K_{m+1, m+2, \dots, n}|1}
\gb{n|K_{m+1, m+2, \dots, n}|m}},}}
where
$s$ and $t$ are given by eq.~\atwentynine\ and u is given by
\eqn\athirtyfour{ u=(K_{m+1, m+2, \dots,
n})^2=(K_{m+1}+K_{m+2}+\dots +K_{n})^2.}
%


\newsec{Application to Three and Higher Loops}


Ideas presented in the previous sections can be applied to higher
loops. Let us consider triple-box configurations appearing at
three-loops. The configurations we consider are obtained from the
double-box configurations at two loops by adding three new
propagators to form the third loop. This way, one can produce a
ladder diagram as in fig. 19a or a double box with a pentagon as
in fig. 19b. We make a slight abuse of terminology and call both
kind of configurations triple-box diagrams.
\ifig\figthreeone{Three-loop triple box configurations. $(a)$ A
triple box ladder integral. $(b)$ A double box with a pentagon.}
{\epsfxsize=0.95\hsize\epsfbox{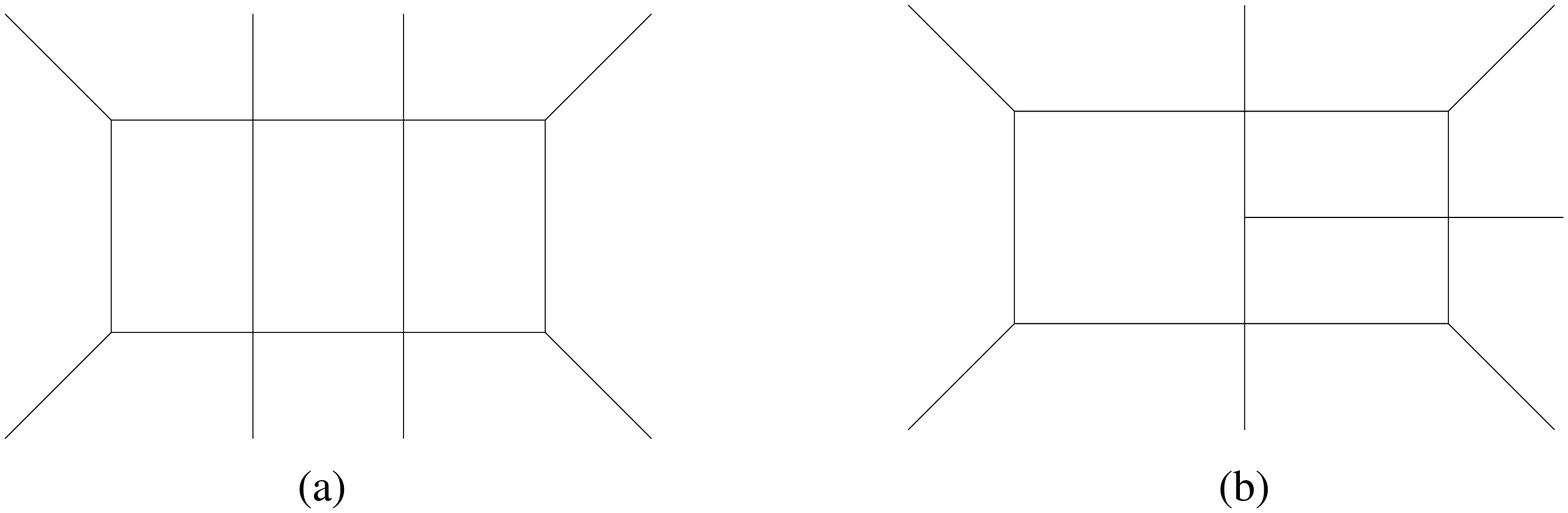}}
\noindent Every triple box contains ten propagators\foot{Of
course, for a large enough number of gluons one can also find
split triple boxes which can have 11 or 12 propagators.}.
Therefore it is natural to start with a ten-particle cut. This
produces ten delta-functions whereas the number of integration
variables is twelve. However, it follows from our previous
analysis that box configurations develop additional
propagator-like singularities which can also be cut (replaced by
their discontinuities). A triple-box configuration naturally
admits two extra propagator-type singularities which allows us to
consider twelve-particle cuts. Therefore, it should be possible to
completely localize all momentum integrals, at least if the number
of gluons is not big enough. Then it is straightforward to write
down an expression for the coefficients analogous to eq.~\aseven.
Obviously, a similar analysis can be performed at any number of
loops. It is interesting to mention that at three loops, some of
the triple boxes that enter in the calculation of the four-gluon
amplitude are not scalar triple boxes \Bernry. This means that the
numerator in the integrand is not one but an inverse propagator.
See fig. 20. This was also found to be the case for higher loops
\Bernry.

We start our discussion with the analysis of ladder diagrams which
allow a straightforward generalization of our discussion in
section 4. Then we turn to the triple-box integrals where one of
the ``boxes" has five propagators and see how our analysis of
singularities realizes the phenomenon mentioned above. For
concreteness, we will concentrate on the four-gluon amplitude
though an identical analysis can be performed regardless of the
number of external lines. Let us start with the triple-box
configuration in fig. 21.
\ifig\figthreetwo{Schematic representation of a modified ``triple
box" integral given in \Bernry.}
{\epsfxsize=0.50\hsize\epsfbox{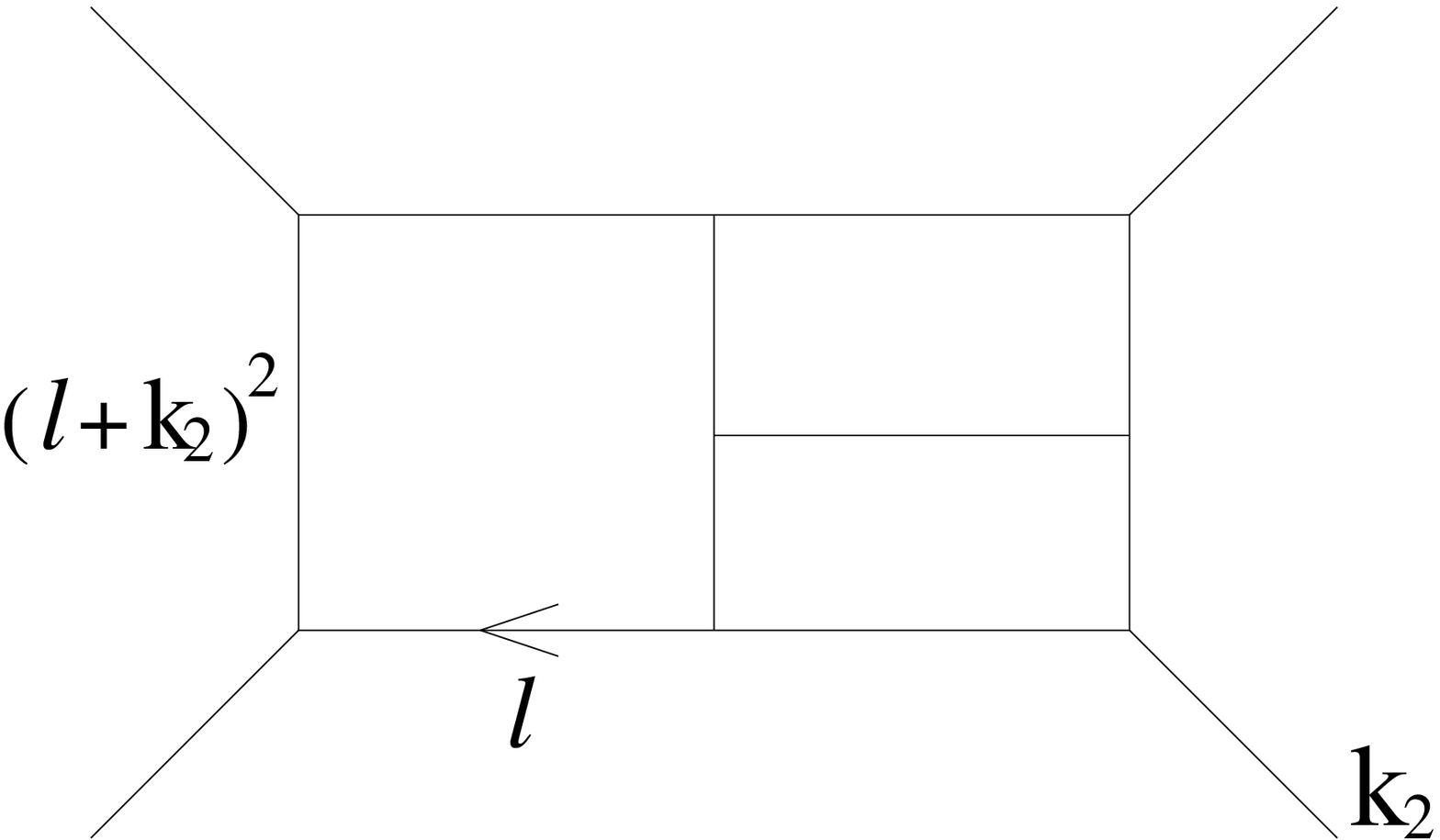}}
\noindent
\ifig\figthreethree{A ladder triple-box configuration with four
external gluons.} {\epsfxsize=0.50\hsize\epsfbox{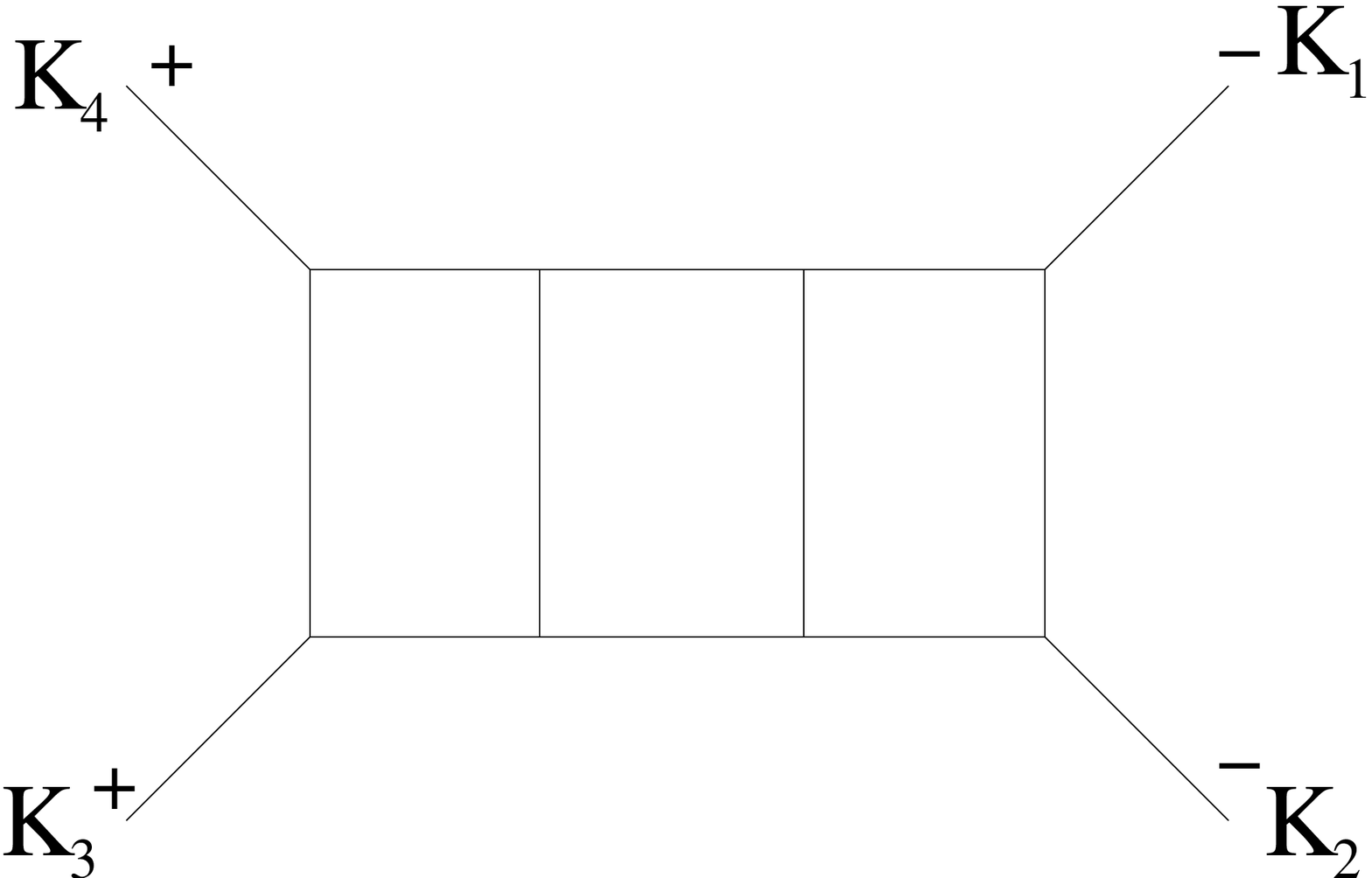}}
\noindent
\ifig\figthreefour{One of the helicity configurations contributing
to the coefficient of the ladder triple-box in fig.21.}
{\epsfxsize=0.50\hsize\epsfbox{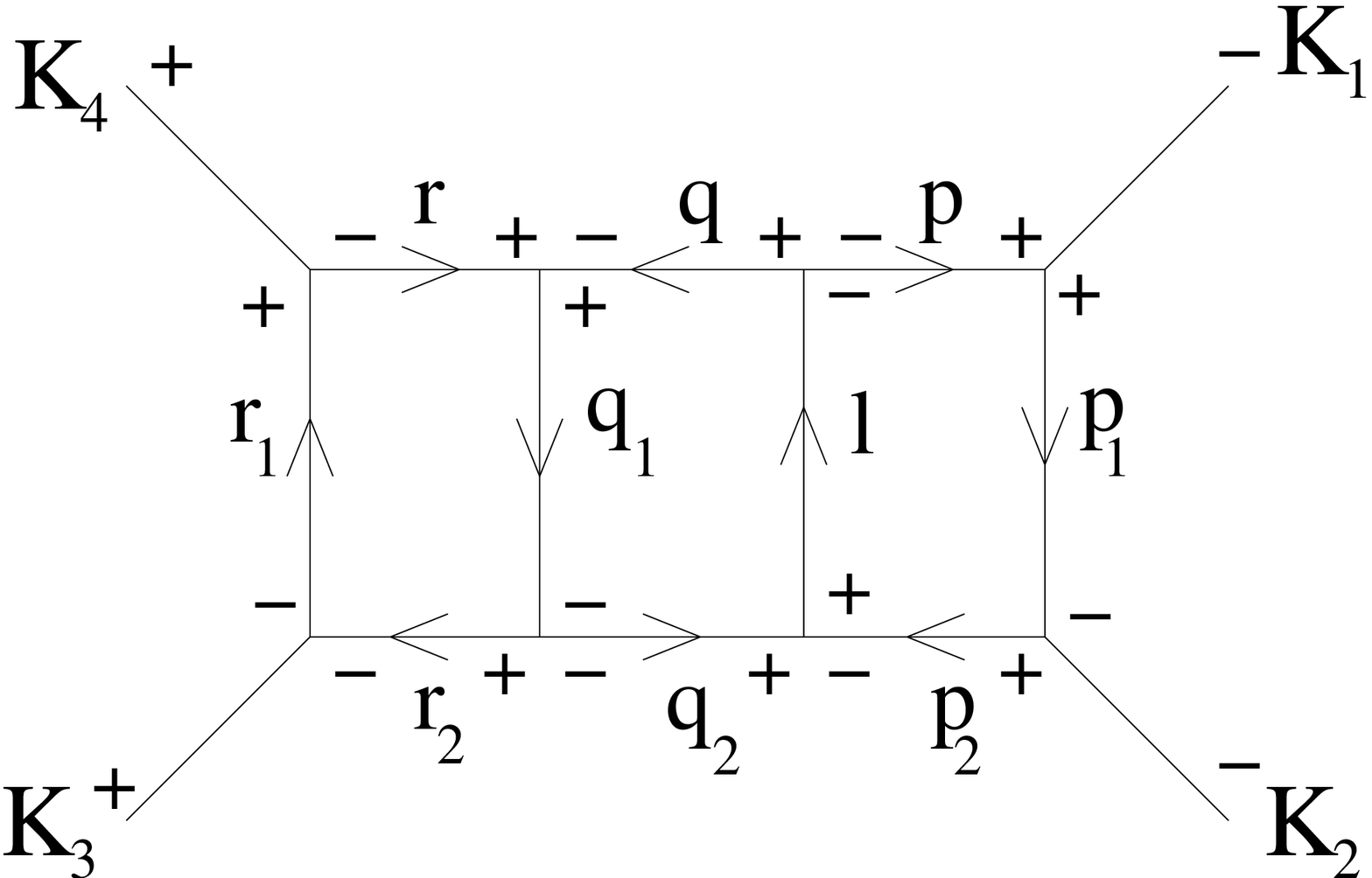}}
\noindent
In this case, there are twelve helicity configurations, all giving the same
answer.
It is enough to consider one of them, for example the one
in~\figthreefour.
Similarly to the two-loop case, it is enough to consider the ten-particle cut because
the measure integral factors out and cancels. Then the corresponding coefficient is
given by
\eqn\bone{
d_1=-i^3 \tilde{A}_4^{tree}s^2 \tilde{t}^2
{\vev{r_1~r_2}^3 \over \vev{r_2~3} \vev{3~r_1}}
{[r_1~4]^3 \over [4~r] [r~r_1]},}
where $\tilde{A}_4^{tree}$ is the tree level four-gluon amplitude
with the external lines $(K_1^-, K_2^-, r_2^+, r_1^+)$,
$s=(K_1+K_2)^2$ and $\tilde{t}= (K_1-r)^2$. In eq.~\bone, we used
that the product of the six tree amplitudes was computed in
section 3. Using momentum conservation, eq.~\bone\ can be
simplified to give
\eqn\btwo{ d_1= -iA_4^{tree}s^3t.}
where $t=(K_2+K_3)^2$. This coincides with the answer
from~\Bernry.
\ifig\figthreefive{A triple-box configuration.}
{\epsfxsize=0.50\hsize\epsfbox{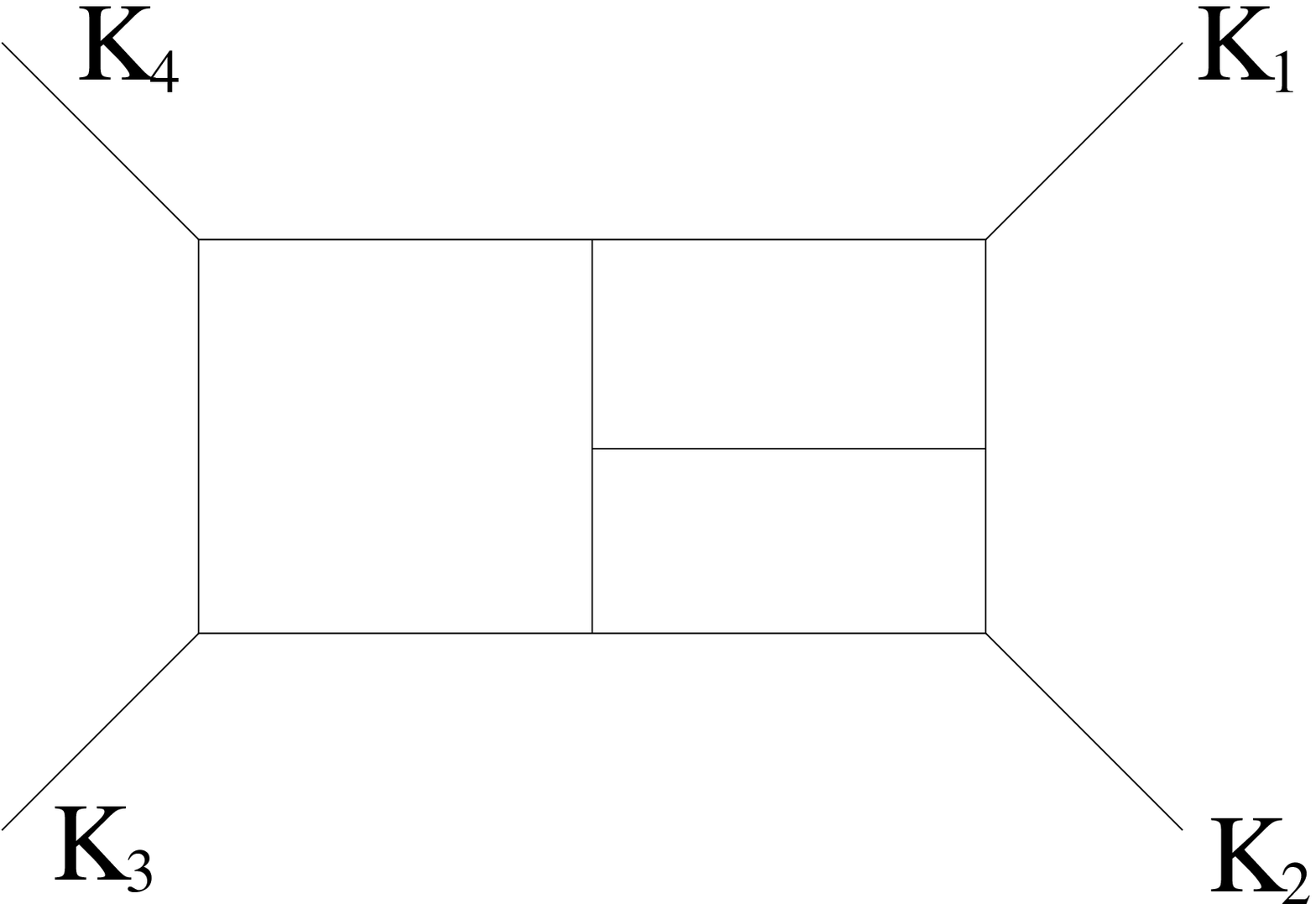}}
\noindent Now let us consider the configuration in~\figthreefive.
Note that one of the loops has five propagator and this is why we
said that the configuration was a double box with a pentagon. This
is the basic reason why the integral is not a scalar box integral
as we will see below. Let us start our analysis with the measure
integral
\eqn\bthree{\eqalign{ {\cal I}_0= &\int d^4l d^4q d^4p \delta
(l^2) \delta((l-K_3)^2) \delta ((l-K_3-K_4)^2)
\delta((l-K_3-K_4-K_1)^2) \cr & \delta((p-K_2-l)^2) \delta
((p-K_2)^2) \delta((q-K_1)^2) \delta ((p+q)^2) \delta (p^2) \delta
(q^2)}}
and perform the integration over $p$ and $q$. After integrating
over $p$ we obtain (up to the momenta-independent factor)
\eqn\bfour{
{1 \over (K_2+q)^2 (K_2+l)^2}.}
Then we cut the ``propagator'' $1/(K_2+q)^2$. This gives the
fourth delta-function which allows us to perform the integration
over $q$. This produces (again, we ignore the momenta-independent
factors) one more factor of $(K_2+l)^2$ in the denominator. Thus,
we obtain that this triple box configuration has a singularity
\eqn\newsin{\left( {1\over (K_2+l)^2}\right)^2 .}
Let us calculate the coefficient of this triple box integral by
calculating the product of the eight-gluon amplitudes. For every
non-vanishing helicity configuration, the product of the six gluon
amplitudes on the right is the coefficient $c_t$ of the two-loop
four-gluon amplitude studied in section 3 and it is given by
\eqn\bfive{
-A_4^{\prime tree}st^{\prime 2},}
where $A_4^{\prime tree}$ is the tree four-gluon amplitude with
external lines $(K_1^-, K_2^-, l^+, (l-K_3-K_4)^+)$ and $t^{\prime}$
is given by
\eqn\bsix{
t^{\prime}=(K_2+l)^2.}
Therefore, in attempting to calculate the integral of the product
of the tree amplitudes, the singularity \newsin\ cancels out. This
means that the coefficient of the triple box
in~\figthreefive\ is zero. This is in agreement with results
of~\Bernry. On the other hand, the amplitude $A_4^{\prime tree}$
has a factor
\eqn\bseven{
{1 \over \vev{2~l}}={[2~l] \over (K_2+l)^2}.}
This indicates that the actual diagram has a singularity ${1 \over
(K_2+l)^2}$. In order to account for this singularity we have to
introduce a slightly modified triple box integral schematically
shown in~\figthreetwo. The basic idea is to multiply in the
numerator by $(K_2+l)^2$ in order to cancel one of the power in
\newsin\ and get the correct $1/(K_2+l)^2$ singular behavior. This shows that this
triple-box integral should be in the list of scalar integrals of
the amplitude under study. This is completely consistent with
results of~\Bernry. Now we can cut this ``propagator'' and
completely localize the integral. Note that the combination
\eqn\beight{
[2~l] \delta ((K_2+l)^2)}
is not necessarily zero. This just means that we have to choose
the solution
\eqn\bnine{
\lambda_l \sim \lambda_2.}
The coefficient of this modified box
is now straightforward to compute. The answer is
\eqn\bten{
d_2=-iA_{4}^{tree}s^2t.}
This coincides with the corresponding coefficient from~\Bernry.
All other coefficients of this amplitude can be found in a similar manner.
Even though we concentrated on the four-gluon amplitude, we
could do the same analysis for any amplitude admitting additional cuts.

As mentioned in the introduction, the basis of integral for $\N=4$
$L$-loop amplitudes of gluons is not known except for $L=1$. One
can imagine that a more systematic analysis along the lines of the
discussion presented in this section might give a way of obtaining
such a basis. It would be interesting to explore this direction in
the future.


\newsec{Conclusion}


In this paper, we observed that certain scalar double-box integrals
which appear at two loops in $\N = 4$ Yang-Mills theory possess
hidden singularities. Such singularities are manifest after a
quadruple cut is performed on one of the boxes. The end result is
that one can straightforwardly calculate the coefficient of such
integrals by an octa-cut which localizes the cut integral. The form
of the coefficient is universal and it is given by the product of a
certain number of tree-level amplitudes. This technique is
applicable to all scalar double box integrals in amplitudes with
less than seven external gluons and to a large subset of double box
integrals for seven or more external gluons. The basis of integrals
at two loops is not known in general. For four gluons the amplitude
is given in terms of only scalar double-box integrals. If it turns
out that the basis of integrals for five- and six-gluon amplitudes
is also given by scalar double box integrals, then our technique
gives a simple way of computing all those amplitudes for any
helicity configuration. We also argued that this technique can be
applied to higher loop amplitudes. At three loops we found that our
technique can be easily extended to compute the coefficient of
ladder diagrams. For the class of diagrams with a pentagon, our
method shows that the coefficient of scalar integrals is zero and
naturally gives the modified integral for which the coefficient does
not vanish.


\bigskip

\centerline{\bf Acknowledgments}

\bigskip


The authors would like to thank Z. Bern, R. Britto, L. Dixon, B. Feng, H. Osborn, M. Spradlin
and A. Volovich for comments on the first version of the manuscript.
Research of E.I.B is supported by NSF grant PHY-0070928. Research
of F.C. is supported in part by the Martin A. and Helen Chooljian
Membership at the Institute for Advanced Study and by DOE grant
DE-FG02-90ER40542. Research at the Perimeter Institute is
supported by funds from NSERC of Canada.


\listrefs

\end